\documentclass[aip,jcp,reprint,eqsecnum,superscriptaddress,twocolumn,longbibliography,floatfix]{revtex4-2}

\usepackage{empheq}
\usepackage{graphicx}
\usepackage{dcolumn}
\usepackage{bm}

\usepackage[utf8]{inputenc}
\usepackage[T1]{fontenc}
\usepackage{mathptmx}
\usepackage{etoolbox}
\usepackage{xcolor}
\usepackage{mathpazo}
\usepackage{bm}
\usepackage[unicode=true,
 bookmarks=true,bookmarksnumbered=false,bookmarksopen=false,
 breaklinks=false,pdfborder={0 0 0},pdfborderstyle={},backref=false,colorlinks=true]
 {hyperref}
\hypersetup{pdfborderstyle={},pdfborderstyle={},pdfborderstyle={},pdfborderstyle={},pdfborderstyle={},pdfborderstyle={},linkcolor=blue,citecolor=blue}
\usepackage{breakurl}

\usepackage{natbib}
\usepackage{amssymb,amsmath}

\newcommand{\bp}{\beta p}
\newcommand{\bpc}{\beta p_{\text{cr}}}
\newcommand{\pc}{p_{\text{cr}}}
\newcommand{\lambdac}{\lambda_{\text{cr}}}
\newcommand{\cl}{\text{cp}}
\newcommand{\dd}{\mathrm{d}}

\newcommand\beq{\begin{equation}}
\newcommand\eeq{\end{equation}}
\newcommand\beqa{\begin{eqnarray}}
\newcommand\eeqa{\end{eqnarray}}
\newcommand{\nn}{\nonumber\\}
\def\bal#1\eal{\begin{align}#1\end{align}}

\makeatletter
\def\@email#1#2{%
 \endgroup
 \patchcmd{\titleblock@produce}
  {\frontmatter@RRAPformat}
  {\frontmatter@RRAPformat{\produce@RRAP{*#1\href{mailto:#2}{#2}}}\frontmatter@RRAPformat}
  {}{}
}%
\makeatother
\begin{document}

\title[]{Structural properties of hard-disk fluids under single-file confinement}
\author{Ana M. Montero}
\affiliation{Departamento de F\'isica, Universidad de Extremadura, E-06006 Badajoz, Spain}
\email{anamontero@unex.es}
\author{Andr\'es Santos}%
 \affiliation{
Departamento de F\'isica, Universidad de Extremadura, E-06006 Badajoz, Spain
and Instituto de Computaci\'on Cient\'ifica Avanzada (ICCAEx), Universidad de Extremadura, E-06006 Badajoz, Spain
}%
\email{andres@unex.es}

\date{\today}

\begin{abstract}
The structural properties of confined single-file hard-disk fluids are studied analytically by means of a mapping of the original system onto a one-dimensional mixture of non-additive hard rods, the mapping being exact in the polydisperse limit. Standard statistical-mechanical results  are used as a starting point to derive thermodynamic and structural properties of the one-dimensional mixture, where the condition that all particles have the same chemical potential must be taken into account. Analytical results are then provided for the $n$th neighbor probability distribution function, the radial distribution function, and the structure factor, a very good agreement being observed upon comparison with simulation data from the literature. Moreover, we have analyzed the scaling form for the disappearance of defects in the zigzag configuration for high pressure, and have obtained the translational correlation length and the structural crossover in the oscillation frequency for asymptotically large distances.
\end{abstract}

\maketitle

\section{\label{sec:introduction}Introduction}
The study of the structural properties of any given liquid system is a key step in completely understanding its behavior and the nature of the spatial correlations induced by the interactions between its particles.\cite{BH76,HM13,S16} These structural properties go beyond the purely thermodynamic ones and provide insight into the arrangement and behavior of the particles of the system.\cite{BPGS06,ONCB06,RHS07,BSB11,MS16,SYH20} Among these properties, the radial distribution function (RDF) and the structure factor are two of the most relevant ones, the former because it describes how the local density of particles varies with distance from a reference particle, and the latter due to its direct connection with diffraction experiments.

Despite its clear importance, systems whose structural properties are amenable to exact analytic solutions are very scarce, and usually limited to one-dimensional (1D) systems with only nearest-neighbor interactions.\cite{T36,KT68,HC04,S07b,S07,VG11,S16,FS17,MS19,MS20,FMS21} Otherwise, one must resort to approximations, numerical methods, or simulations.

Highly confined two- and three-dimensional systems, where the available space along one of the dimensions of the pore is much larger than along the other ones, in such a way that particles are confined into single-file formation,\cite{B62,B64b,WPM82,PK92,KP93,P02,KMP04,FMP04,VBG11,GV13,GM14,M14b,M15,GM15,HFC18,M20,HBPT20,P20,PBT22,JF22} make an interesting and special class of systems. Their most relevant properties are the longitudinal ones, and they can be studied by treating the system as  quasi one-dimensional (Q1D). These properties are often amenable to an exact statistical-mechanical solution,\cite{B64b,KP93,GV13,MS23} which makes Q1D systems a particularly relevant field of study, especially since, despite their simplicity, they can be used to gain valuable insight into phenomena occurring in real confined fluids.

The Q1D hard-disk fluid belongs to this last class of systems, and its study is an active field of research\cite{P20,PBT22,ZGM20,HBPT20,M20,HBPT21,HC21,MS23} due to a combination of a manageable interaction potential and a large variety of situations it can be applied to. However, even under these favorable circumstances, structural properties of the Q1D hard-disk fluid are problematic to obtain from the transfer-matrix method,\cite{HFC18,GV13,GM15,RGM16,PBT22} and thus, they are usually studied by means of simulations\cite{VBG11,HBPT21} or the so-called planting method,\cite{HC21} which also requires averaging over randomly generated configurations.

In this paper, we take a somewhat different approach by exploiting a mapping of the original Q1D system onto a 1D polydisperse mixture of \emph{non-additive} hard rods. The peculiarity of the mapped mixture is that since all of its 1D \emph{species} actually represent the same type of disk,  the condition that all {species} of the mixture have the same chemical potential must be taken into account.
Standard liquid theory of mixtures\cite{S16} is used on the newly mapped 1D mixture to compute the structural properties of the original Q1D system. To obtain explicit results, we take \emph{discrete} mixtures with a large, but finite, number of species. In that way, the exact properties of the Q1D fluid are recovered by taking the \emph{continuous} polydisperse limit.

Our paper is organized as follows. Section~\ref{sec:solution} defines the system under study and its main properties. Section~\ref{sec:3} presents theoretical results regarding the thermodynamic and structural properties of generic 1D mixtures with nearest-neighbor interactions. This theoretical background is subsequently used in  Sec.~\ref{sec:4}, which contains an analysis of the results obtained for the neighbor probability distribution functions, the RDF, and the structure factor.
In addition,  the disappearance of defects in the zigzag configuration for high pressure is analyzed. Moreover, the asymptotic behavior for large distances is studied by identifying the translational correlation length and a structural crossover in the oscillation frequency.
Finally, Sec.~\ref{sec:conclusions} closes the paper with a presentation of the main conclusions.

\section{\label{sec:solution}The system}

\subsection{\label{sec:systemDefinition}Q1D hard-disk fluid}

Consider a  two-dimensional system of $N$ hard disks interacting via a pairwise potential of the form
\begin{equation}
    \varphi(r) = \begin{cases} \infty &\text{if } r<1 \\ 0 &\text{if } r>1\end{cases},
\end{equation}
where, for simplicity, the hard-core diameter of the particles is assumed to be equal to $1$. The particles are confined in a very long channel of width $w = 1+\epsilon$ and length $L \gg w$, in such a way that they are in single-file formation, and only first nearest-neighbor interactions take place. These two conditions set the range of validity of the excess pore width to  $0\leq \epsilon \leq \epsilon_{\mathrm{max}}$, where
$\epsilon_{\mathrm{max}}=\sqrt{3}/2\simeq 0.866$.
Note that, if the transverse separation between two disks at contact is $\Delta y$, their longitudinal separation is then
\beq
\label{eq:a(s)}
a(\Delta y)\equiv\sqrt{1-(\Delta y)^2}.
\eeq

Due to the highly anisotropic nature of this confined system, it is often useful to characterize it via its longitudinal properties, such as the number of particles per unit length, $\lambda \equiv N/L$, or the reduced pressure $p \equiv P_{\|} \epsilon$, where $P_{\|}$ is the longitudinal component of the pressure.
At close packing, the linear density reaches a maximum value of $\lambda_{\mathrm{cp}}(\epsilon) = 1/a(\epsilon)$, and the reduced pressure diverges.

From the exact transfer-matrix solution of this Q1D system,\cite{KP93} one can obtain the equation of state as
\bal
\label{eq:z_exact_01}
    Z \equiv \frac{\bp}{\lambda}
    = &  1 + \frac{\bp}{\ell} \int_{-\frac{\epsilon}{2}}^{\frac{\epsilon}{2}}\mathrm{d}y\int_{-\frac{\epsilon}{2}}^{\frac{\epsilon}{2}}\mathrm{d}y' \,e^{-\bp a(y-y')}a(y-y')\nn
    &\times\phi(y)\phi(y'),
\eal
where $\beta\equiv 1/k_BT$ ($k_B$ and $T$ being the Boltzmann constant and the absolute temperature, respectively), $\ell$ is the maximum eigenvalue of the problem
\begin{equation}
\label{eq:2.5}
\int_{-\frac{\epsilon}{2}}^{\frac{\epsilon}{2}}\mathrm{d}y'\, e^{-\bp a(y-y')}\phi(y')=\ell \phi(y),
\end{equation}
and $\phi(y)$ is the associated eigenfunction. Moreover, $\phi^2(y)$ is the probability density profile along the transverse direction $y$.
An expression for the isothermal susceptibility  $\chi_T\equiv\beta^{-1}\partial_p\lambda$ is derived in Appendix~\ref{app0}. This quantity has been recently seen to encode  how dynamic correlations  in transient one-dimensional diffusive systems depend on spatial fluctuations of the initial state.\cite{BJC22}

In a recent study,\cite{MS23} we derived the exact third and fourth virial coefficients from Eq.~\eqref{eq:z_exact_01} and proved that near close packing, $Z\to 2/(1-\lambda/\lambda_\cl)$.
Additionally, as a practical alternative to the numerical solution of Eq.~\eqref{eq:2.5}, we proposed two approximate transverse profiles: a simple uniform profile, $\phi(y)\to\text{const}$, and a more sophisticated  exponential-like profile, $\phi(y) \to e^{-\bp a(y+\frac{\epsilon}{2})}+e^{-\bp a(y-\frac{\epsilon}{2})}$. Comparison with transfer-matrix and simulation results showed a good performance of both approximations, especially the quasi-exponential one.

As said in Sec.~\ref{sec:introduction}, in this work, we focus on the longitudinal structural properties of the confined hard-disk fluid by taking advantage of its mapping  onto a 1D mixture of non-additive hard rods (see Appendix A of Ref.~\onlinecite{MS23}).

Let us first introduce the RDF of the confined fluid. The local number density is $n_1(\mathbf{r})=\lambda\phi^2(y)$ and the two-body distribution function is $n_2(\mathbf{r},\mathbf{r}')=n_1(\mathbf{r})n_1(\mathbf{r}')g(\mathbf{r},\mathbf{r}')$, where $g(\mathbf{r},\mathbf{r}')$ is the RDF. For simplicity, we keep the term ``radial,'' although in contrast to isotropic fluids, $g(\mathbf{r},\mathbf{r}')$ is not a function of $|\mathbf{r}-\mathbf{r}'|$ only, but depends on $y$, $y'$, and $|x-x'|$. To make that more explicit, we introduce the changes of notation $n_1(\mathbf{r})\to n_1(y)$, $n_2(\mathbf{r},\mathbf{r}')\to n_2(y,y';|x-x'|)$, and $g(\mathbf{r},\mathbf{r}')\to g(y,y';|x-x'|)$.

\subsection{1D hard-rod mixture}
\label{sec:2B}

The mapping is based on the idea that the transverse coordinate of each disk, $-\epsilon/2 < y < \epsilon/2$, represents the dispersity parameter of the mixture, and therefore, each species  in the hard-rod mixture maps the transverse coordinate of the original Q1D system. Since $y$ is a continuous variable, the equivalent 1D mixture has a continuous distribution of components. In practice, however, it is enough to take a discrete mixture with a sufficiently large number $M$ of components to accurately describe the system, as will be shown in Sec.~\ref{sec:result_finiteM}.

\begin{figure}
    \includegraphics[trim={1cm 1cm 2cm 0.5cm},clip,width=0.95\columnwidth]{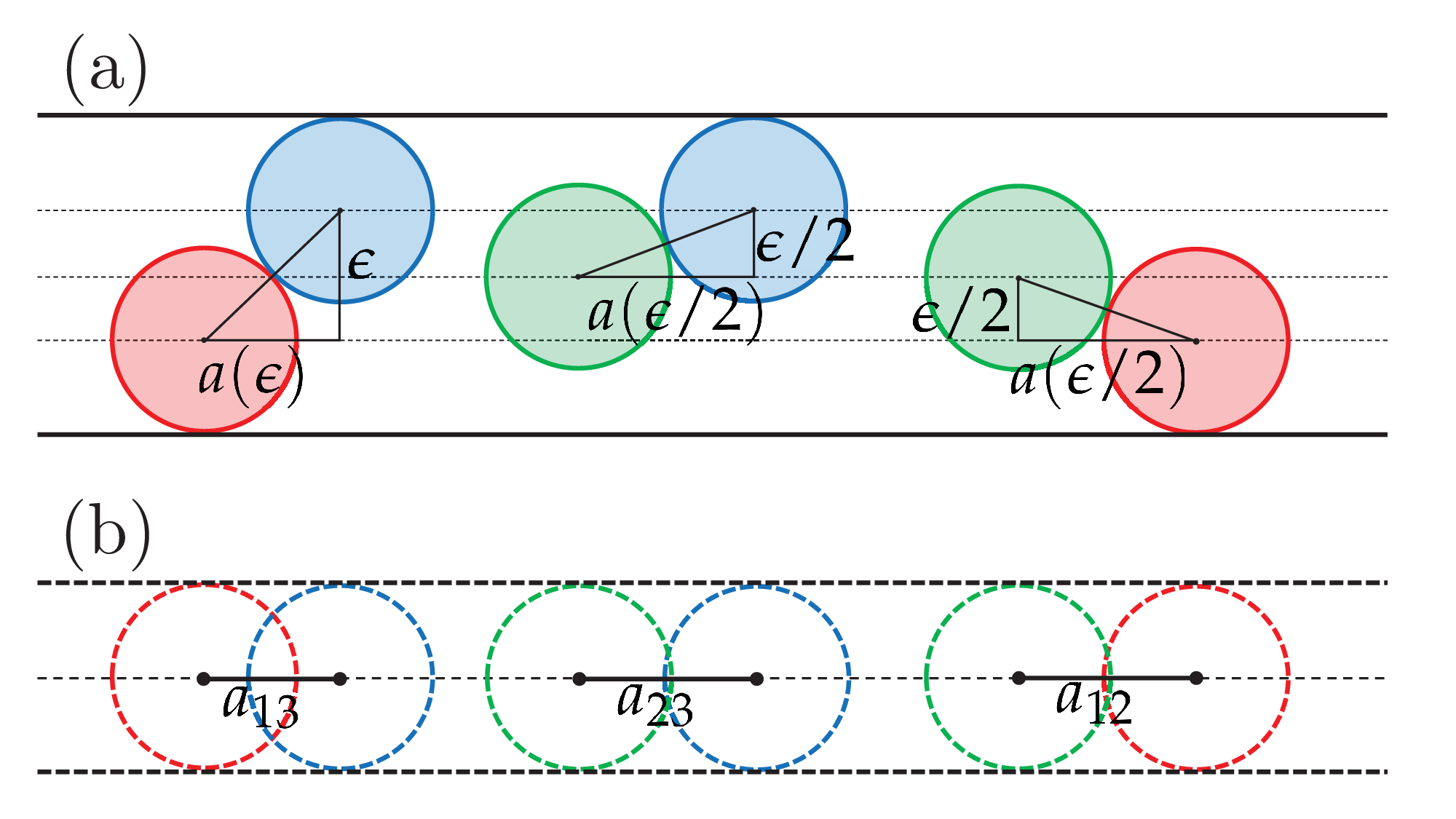}\\
    \caption{Schematic representation of the mapping of (a) the original Q1D system onto (b) a 1D mixture of non-additive hard rods. In this illustration,  the number of species chosen in the mapping has been set to $M=3$ for simplicity. Note that $a_{11}=a_{22}=a_{33}=1$, but $a_{13}=a(\epsilon)<a_{12}=a_{23}=a(\frac{\epsilon}{2})<1$.}
    \label{fig:Sketch}
\end{figure}

Under this framework, each species $i$ of a discrete $M$-component mixture represents a disk whose vertical coordinate is
\begin{equation}
y_i=-\frac{\epsilon}{2}+(i-1)\delta y,\quad i=1,2,\ldots,M,\quad \delta y\equiv \frac{\epsilon}{M-1}.
\end{equation}
The hard-core distance between two rods of species $i$ and $j$ is equal to the longitudinal distance at contact of the two disks they represent, i.e.,
\begin{equation}
a_{ij}=a(y_i-y_j)=\sqrt{1-\left[(i-j)\delta y\right]^2}.
\end{equation}
Note that $a_{ii}=1$ but $a_{ij}<1$ if $i\neq j$, so that the hard-rod mixture is negatively non-additive. Figure~\ref{fig:Sketch} shows a schematic representation of this mapping with $M=3$.

Before applying this 1D mapping to obtain the (longitudinal) structural properties of the original Q1D fluid, let us present the main properties of a generic 1D mixture of particles with nearest-neighbor interactions.

\section{1D mixtures  with nearest-neighbor interactions}
\label{sec:3}

\subsection{Spatial correlations}

Let us consider an $M$-component 1D mixture made of $N$ particles ($N_i$ belonging to species $i$) with a linear number density $\lambda$. The interaction potential between two particles of species $i$ and $j$, $\varphi_{ij}(x)$, is assumed to act only if those particles are nearest neighbors.

The key quantities are the  probability density distributions, $P_{ij}^{(n)}(x)$, such that  $P_{ij}^{(n)}(x)\dd x$ is the probability that the $n$th neighbor of a reference particle of species $i$ belongs to species $j$ and is located at a distance between $x$ and $x+\dd x$ from the reference particle.
Note that $P_{ij}^{(n)}(x)\neq P_{ji}^{(n)}(x)$ but $x_i P_{ij}^{(n)}(x)=x_j P_{ji}^{(n)}(x)$, where $x_i=N_i/N$ denotes the mole fraction of species $i$.
The total $n$th neighbor probability distribution function is defined as
\begin{equation}\label{eq:pn_def}
    P^{(n)}(x) = \sum_{i,j} x_i P^{(n)}_{ij}(x).
\end{equation}
Then, the partial and total RDF are  given by
\begin{subequations}
\beq
\label{3.0}
g_{ij}(x)=\frac{1}{\lambda x_j}\sum_{n=1}^\infty P_{ij}^{(n)}(x),
\eeq
\bal
\label{eq:g_def}
g(x)=&\sum_{i,j}x_i x_j g_{ij}(x)=\frac{1}{\lambda}\sum_{n=1}^\infty P^{(n)}(x).
\eal
\end{subequations}
The structure factor, $S(q)$, is directly related to the Fourier transform of the total correlation function $h(x)\equiv g(x)-1$,
\bal
\label{3.01}
S(q)=&1+\lambda \int_{-\infty}^\infty \dd x\, e^{-\imath q x}h(x)\nn
=&1+2\lambda \int_{0}^\infty \dd x\, \cos( q x)h(x),
\eal
where $\imath$ is the imaginary unit.

From standard statistical-mechanical results in the isothermal-isobaric ensemble, one finds\cite{S16}
\begin{subequations}
\beq
\label{3.1a}
P_{ij}^{(1)}(x)=\sqrt{\frac{x_j}{x_i}} A_i A_je^{-\beta\left[\varphi_{ij}(x)+p x\right]},
\eeq
\beq
\label{3.1b}
P_{ij}^{(n)}(x)=\sum_k\int_0^x \dd x'\,P_{ik}^{(n-1)}(x')P_{kj}^{(1)}(x-x').
\eeq
\end{subequations}
In Eq.~\eqref{3.1a},  the parameters $\{A_i\}$ are given by the solution of the nonlinear set of equations,\cite{note_23_04_1}
\beq
\label{eq:solve01}
 A_i \sum_j\Omega_{ij}(\beta p)\sqrt{x_j} A_j=\sqrt{x_i},
\eeq
where
\begin{equation}
    \Omega_{ij}(s) = \int_0^\infty \mathrm{d}x\, e^{-sx}e^{-\beta \varphi_{ij}(x)}
\end{equation}
is the Laplace transform of the Boltzmann factor.
Notice that, for simplicity, we omit in the notation the dependence of $\Omega_{ij}(s)$ on $\beta$. The physical condition $\lim_{x\to\infty}\varphi_{ij}(x)=0$ implies that $\lim_{s\to 0} s \Omega_{ij}(s)=1$. As a consequence, from Eq.~\eqref{eq:solve01}, we have $\lim_{p\to 0}A_i/\sqrt{\bp x_i}=1$.

The convolution structure of Eq.~\eqref{3.1b} suggests the introduction of the Laplace transforms $\widetilde{P}_{ij}^{(n)}(s)$, $\widetilde{G}_{ij}(s)$, and $\widetilde{G}(s)$ of ${P}_{ij}^{(n)}(x)$, $g_{ij}(x)$, and $g(x)$, respectively, so that
\begin{subequations}
\beq
\label{3.5a}
\widetilde{P}_{ij}^{(1)}(s)=\sqrt{\frac{x_j}{x_i}} A_i A_j\Omega_{ij}(s+\bp),
\eeq
\beq
\label{3.5b}
\widetilde{P}_{ij}^{(n)}(s)=\left(\left[\widetilde{\mathsf{P}}^{(1)}(s)\right]^n\right)_{ij},
\eeq
\bal
\label{3.5c}
\widetilde{G}_{ij}(s)=&\frac{1}{\lambda x_j}\left(\sum_{n=1}^\infty\left[\widetilde{\mathsf{P}}^{(1)}(s)\right]^n\right)_{ij}\nn
=&\frac{1}{\lambda x_j}\left(\widetilde{\mathsf{P}}^{(1)}(s)\cdot\left[\mathsf{I}-\widetilde{\mathsf{P}}^{(1)}(s)\right]^{-1}\right)_{ij},
\eal
\beq
\label{3.5d}
\widetilde{G}(s)=\sum_{i,j} x_i x_j \widetilde{G}_{ij}(s).
\eeq
\end{subequations}
Here, $\widetilde{\mathsf{P}}^{(1)}(s)$ is the $M\times M$ matrix of elements $\widetilde{P}^{(1)}_{ij}(s)$, and $\mathsf{I}$ is the corresponding unit matrix.
Notice that Eq.~\eqref{3.5c} can be rewritten as
\beq
\label{B1}
\frac{1}{\lambda}\widetilde{P}_{ij}^{(1)}(s)=x_j\widetilde{G}_{ij}(s)-\sum_k x_k\widetilde{G}_{ik}(s)\widetilde{P}_{kj}^{(1)}(s).
\eeq
In turn, the structure function defined by Eq.~\eqref{3.01} can be obtained from $\widetilde{G}(s)$ as
\bal
\label{eq:structure_factor}
{S}(q) = 1 + \lambda \left[ \widetilde{G}(s)+\widetilde{G}(-s)\right]_{s=\imath q}.
\eal

\subsection{Thermodynamic quantities. Physical meaning of the parameters $\{A_i\}$}
From the physical condition $\lim_{x\to\infty}g_{ij}(x)=1$, one finds the equation of state (see Appendix~\ref{appA})
\beq
\label{eq:number_density}
\frac{\beta}{\lambda}=-\sum_{i,j}\sqrt{x_i x_j}A_i A_j\partial_p\Omega_{ij}(\beta p).
\eeq
In order to derive the Gibbs free energy $G$, we need to rewrite Eq.~\eqref{eq:number_density} in an alternative form. First, taking into account from Eq.~\eqref{eq:solve01} that $\partial_p \sum_{i,j}\sqrt{x_i x_j}A_i A_j\Omega_{ij}(\beta p)=\partial_p\sum_i x_i=0$, one has $\beta/\lambda=2\sum_{i,j}\sqrt{x_i x_j}A_j\Omega_{ij}(\beta p)\partial_pA_i$. Second, using again Eq.~\eqref{eq:solve01}, $\beta/\lambda=2\sum_i {x_i}A_i^{-1}\partial_pA_i$. Therefore,
\beq
\label{eq:number_density1}
\frac{\beta}{\lambda}=\sum_i x_i \partial_p \ln A_i^2.
\eeq
From a practical point of view, Eq.~\eqref{eq:number_density1} is less useful than Eq.~\eqref{eq:number_density} to obtain numerical values since  pressure dependence of $A_i$, in contrast to that of $\Omega_{ij}(\bp)$, is not explicitly known. On the other hand, as we will see, Eq.~\eqref{eq:number_density1} is more compact and convenient at a theoretical level.

By  taking into account Eq.~\eqref{eq:number_density1} in the thermodynamic relation $\lambda^{-1}=N^{-1}(\partial G/\partial_p)_{\beta,\{N_i\}}$, the Gibbs free energy becomes
\beq
\label{eq:Gibbs}
\frac{\beta G}{N}= \sum_i x_i \ln (A_i^2\Lambda_{\text{dB}}),
\eeq
where the integration constant has been determined by the ideal-gas condition $\lim_{p\to 0} \beta G/N=\sum_i x_i\ln(x_i \bp \Lambda_{\text{dB}})$, with $\Lambda_{\text{dB}}\propto \beta^{1/2}$ being the thermal de Broglie wavelength (assumed here to be the same for all species).

Next, we derive the chemical potential $\mu_k=(\partial G/\partial N_k)_{\beta,p,\{N_{i\neq k}\}}$ from Eq.~\eqref{eq:Gibbs},
\beq
\label{eq:chemical_potential}
\beta \mu_k=\ln (A_k^2\Lambda_{\text{dB}})+2\sum_i \frac{N_i}{A_i}\frac{\partial A_i}{\partial N_k}.
\eeq
Differentiating with respect to $N_k$ on both sides of Eq.~\eqref{eq:solve01}, one has
\bal
\frac{N_i}{A_i}\frac{\partial A_i}{\partial N_k}=&\frac{1}{2}\left[\delta_{ik}-\sqrt{\frac{x_i}{x_k}}\Omega_{ik}(\bp)A_iA_k\right]\nn
&-A_i\sqrt{x_i}\sum_j\Omega_{ij}(\bp)\frac{N_j}{\sqrt{x_j}}\frac{\partial A_j}{\partial N_k}.
\eal
Summing over $i$ and applying again Eq.~\eqref{eq:solve01},
\beq
\sum_i \frac{N_i}{A_i}\frac{\partial A_i}{\partial N_k}=\frac{1}{2}(1-1)-\sum_j \frac{N_j}{A_j}\frac{\partial A_j}{\partial N_k},
\eeq
which implies $\sum_i ({N_i}/{A_i})({\partial A_i}/{\partial N_k})=0$. Therefore, Eq.~\eqref{eq:chemical_potential} reduces to
\beq
\label{eq:chemical_condition}
\beta \mu_i=\ln ( A_i^2\Lambda_{\text{dB}}).
\eeq
This provides a physical interpretation of the parameters $\{A_i\}$, namely  $A_i=\sqrt{{z_i}/{\Lambda_{\text{dB}}}}$, where $z_i\equiv e^{\beta \mu_i}$ is the fugacity of species $i$.
To our knowledge, Eqs.~\eqref{eq:number_density1}, \eqref{eq:Gibbs}, and \eqref{eq:chemical_condition} are novel results of the present work.

The internal energy, $U$, can be obtained from $G$ through the thermodynamic relation $U=\left[\partial(\beta G)/\partial\beta\right]_{\bp,\{N_i\}}$. That is,
\beq
\label{eq:internal_energy1}
\frac{\beta U}{N}=\frac{1}{2}+\beta\sum_i {x_i}\left(\frac{\partial \ln A_i^2}{\partial\beta}\right)_{\bp}.
\eeq
Inverting now the steps going from Eq.~\eqref{eq:number_density} to \eqref{eq:number_density1}, except for the change $\bp\leftrightarrow\beta$, we finally have
\beq
\label{eq:internal_energy2}
\frac{\beta U}{N}=\frac{1}{2}-\beta\sum_{i,j}\sqrt{x_i x_j}A_i A_j\left[\frac{\partial \Omega_{ij}(\bp)}{\partial\beta}\right]_{\bp}.
\eeq

\subsection{The equal chemical-potential condition}
\label{sec:3C}
The general  theory of 1D mixtures described above is constructed by taking the mole fractions  $\{x_i\}$ as free thermodynamic variables, independent of $\beta$ and $p$. In general, each species has a distinct chemical potential that, as Eq.~\eqref{eq:chemical_condition} shows, is directly related to the solution of the nonlinear set of equations given by Eq.~\eqref{eq:solve01}.

On the other hand, in the special case of our 1D mixture representing the Q1D fluid, we need to take into account that 1D particles from different species actually represent identical 2D particles  with different transverse coordinates in the original Q1D system, as sketched in Fig.~\ref{fig:Sketch}. This means that the chemical potential of all species must be the same ($\mu_i=\mu$), which implies that all $ A_i= A$ are necessarily also the same. As a consequence, the mole fractions are no longer free variables, but they depend on $\beta$ and $p$, i.e., $\sqrt{x_i}\to \phi_i(\beta,p)$. They are determined by solving Eq.~\eqref{eq:solve01} with $A_i=A$, which now adopts the form of an eigenvalue/eigenvector problem, namely
\beq
\label{eq:eigenproblem}
\sum_j\Omega_{ij}(\beta p)\phi_j=\frac{1}{ A^2}\phi_i.
\eeq

Thus far, in this section we did not need to specify the interaction potentials $\varphi_{ij}(x)$. In the case of the mapped 1D system described in Sec.~\ref{sec:2B}, one simply has $e^{-\beta\varphi_{ij}(x)}=\Theta(x-a_{ij})$, where $\Theta(\cdot)$ is the Heaviside step function, so that $\Omega_{ij}(s)=e^{-s a_{ij}}/s$. Therefore, Eq.~\eqref{eq:eigenproblem} becomes
\beq
\label{eq:eigenproblem2}
\sum_je^{-\bp a_{ij}}\phi_j=\frac{\bp}{ A^2}\phi_i.
\eeq
Moreover, Eq.~\eqref{eq:number_density} yields
\beq
\label{eq:Z}
Z=1+A^2\sum_{i,j}\phi_i\phi_ja_{ij}e^{-\bp a_{ij}}.
\eeq

In what concerns the structural properties, it is proved in Appendix~\ref{app:01} that
\beq
\label{3.19}
P_{ij}^{(n)}(x)=\frac{\phi_j}{\phi_i}A^{2n}Q_{ij}^{(n)}(x),
\eeq
where
\beq \label{eq:qijn_expl}
Q_{ij}^{(n)}(x)=\sum_{k_1}\sum_{k_2}\cdots\sum_{k_{n-1}}R^{(n)}(x;a_{ik_1}+a_{k_1k_2}+\cdots+a_{k_{n-1}j}),
\eeq
with
\beq
\label{eq:qikj}
R^{(n)}(x;\alpha)\equiv\frac{e^{-\beta p x}}{(n-1)!}(x-\alpha)^{n-1}\Theta(x-\alpha).
\eeq
Therefore, the functions $P^{(n)}(x)$ [see Eq.~\eqref{eq:pn_def}], $g_{ij}(x)$ [see Eq.~\eqref{3.0}], and $g(x)$ [see Eq.~\eqref{eq:g_def}] can be expressed as
\begin{subequations}
\beq
P^{(n)}(x)=A^{2n}\sum_{i,j}\phi_i\phi_jQ_{ij}^{(n)}(x),
\eeq
\beq
\label{eq:gx_ij}
g_{ij}(x)=\frac{1}{\lambda\phi_i\phi_j}\sum_{n=1}^\infty A^{2n}Q_{ij}^{(n)}(x),
\eeq
\beq
\label{eq:gx_global}
g(x)=\frac{1}{\lambda}\sum_{n=1}^\infty A^{2n}\sum_{i,j}\phi_i\phi_jQ_{ij}^{(n)}(x).
\eeq
\end{subequations}
Moreover, Eqs.~\eqref{3.5c} and \eqref{B1} become
\begin{subequations}
\beq
\label{3.5cc}
\widetilde{G}_{ij}(s)=\frac{A^2}{\lambda \phi_i\phi_j}\left({\mathsf{\Omega}}(s+\bp)\cdot\left[\mathsf{I}-A^2{\mathsf{\Omega}}(s+\bp)\right]^{-1}\right)_{ij},
\eeq
\beq
\label{B1cc}
\frac{A^2}{\lambda\phi_i}\Omega_{ij}(s+\bp)=\phi_j\widetilde{G}_{ij}(s)-A^2\sum_k \phi_k\widetilde{G}_{ik}(s)\Omega_{kj}(s+\bp),
\eeq
\end{subequations}
where $\mathsf{\Omega}(s)$ is the $M\times M$ matrix with elements $\Omega_{ij}(s)$.

Due to the infinite sum over $n$ in Eqs.~\eqref{eq:gx_ij} and \eqref{eq:gx_global}, one could think that those expressions are merely formal. However, because of the appearance of the Heaviside function in Eq.~\eqref{eq:qikj} and taking into account that $\mathrm{min}\{a_{ij}\}=a(\epsilon)$, the truncation of the sum  at the level of $n=n_{\mathrm{max}}$ yields the exact result up to, at least,  $x\leq n_{\mathrm{max}} a(\epsilon)$.
Alternatively, one can use Eq.~\eqref{3.5cc} to obtain $g_{ij}(x)$ by numerical Laplace inversion.\cite{EulerILT}

It is relevant to note that the knowledge of the partial RDF $g_{ij}(x)$ allows one to obtain not only the longitudinal RDF $g(x)$ but also the two-dimensional RDF $g_{\text{2D}}(r)$, $r=\sqrt{x^2+(\Delta y)^2}$ being the distance between two disks with longitudinal and transverse separations given by $x$ and $\Delta y$, respectively. More specifically, we define
\beq
\label{eq:g_2D}
g_{\text{2D}}(r)=\sum_{i,j}\phi_i^2\phi_j^2 g_{ij}\left(\sqrt{r^2-(y_i-y_j)^2}\right).
\eeq
Quite interestingly, the contact value $g_{\text{2D}}(1^+)$ coincides with the compressibility factor $Z$:
\bal
\label{eq:g_2D_contact}
g_{\text{2D}}(1^+)=&\sum_{i,j}\phi_i^2\phi_j^2 g_{ij}\left(a_{ij}^+\right)
=\frac{A^2}{\lambda}\sum_{i,j}\phi_i\phi_j e^{-\bp a_{ij}}\nn
=&Z,
\eal
where in the last step we have used Eq.~\eqref{eq:eigenproblem2}
\subsection{Continuum limit}
\label{sec:3D}

In the description presented in Secs.~\ref{sec:2B}--\ref{sec:3C}, we have assumed a discrete 1D mixture with a finite (but arbitrary) number of components $M$. In order to fully represent the original Q1D system, where the transverse coordinate $y$ is a continuous variable, one should formally take the continuum limit, $M\to\infty$.
In fact, identifying $\phi_i\to\phi(y_i)\sqrt{\delta y}$,  $A^2\to (\bp/\ell)\delta y$, and taking the limit $M\to\infty$, Eqs.~\eqref{eq:eigenproblem} and \eqref{eq:Z} reduce to Eqs.~\eqref{eq:2.5} and \eqref{eq:z_exact_01}, respectively.

In the continuum case, the role of $P_{ij}^{(n)}(x)$ would be played by $P^{(n)}(y,y';x)$, where $P^{(n)}(y,y';x)\dd y'\dd x$ is the conditional probability that, given  a reference particle with a transverse coordinate $y$, its  $n$th neighbor has a transverse coordinate between $y'$ and $y'+\dd y'$ and  is located at a longitudinal distance between $x$ and $x+\dd x$ from the reference particle. The integral $\int_0^\infty \dd x  \,P^{(1)}(y,y';x)$ is equivalent to the conditional probability defined in Eq.~(6) of Ref.~\onlinecite{HC21}.

The identification $P_{ij}^{(n)}(x)\to P^{(n)}(y_i,y_j;x)\delta y$ allows us to obtain the continuum counterparts of Eqs.~\eqref{eq:pn_def}, \eqref{3.0},  \eqref{eq:g_def}, and \eqref{eq:g_2D} as
\begin{subequations}
\label{3.34abc}
\begin{equation}
\label{3.24a}
    P^{(n)}(x) = \int_{-\frac{\epsilon}{2}}^{\frac{\epsilon}{2}}\dd y \int_{-\frac{\epsilon}{2}}^{\frac{\epsilon}{2}}\dd y'\,\phi^2(y) P^{(n)}(y,y';x),
\end{equation}
\beq
\label{3.24b}
g(y,y';x)=\frac{1}{\lambda \phi^2(y')}\sum_{n=1}^\infty P^{(n)}(y,y';x),
\eeq
\bal
\label{3.24c}
g(x)=&\int_{-\frac{\epsilon}{2}}^{\frac{\epsilon}{2}}\dd y \int_{-\frac{\epsilon}{2}}^{\frac{\epsilon}{2}}\dd y'\,\phi^2(y)\phi^2(y')g(y,y';x)\nn
=&\frac{1}{\lambda}\sum_{n=1}^\infty P^{(n)}(x),
\eal
\bal
\label{eq:g_2D_cont}
g_{\text{2D}}(r)=&\int_{-\frac{\epsilon}{2}}^{\frac{\epsilon}{2}}\dd y \int_{-\frac{\epsilon}{2}}^{\frac{\epsilon}{2}}\dd y'\,\phi^2(y)\phi^2(y')\nn
&\times g\left(y,y';\sqrt{r^2-(y-y')^2}\right).
\eal
\end{subequations}

From Eqs.~\eqref{3.19} to \eqref{eq:qijn_expl}, we conclude that the exact function $P^{(n)}(y,y';x)$ for the Q1D system of single-file hard disks is given by
\beq
\label{3.25}
P^{(n)}(y,y';x)=\frac{\phi(y')}{\phi(y)}\left(\frac{\bp}{\ell}\right)^{n}Q^{(n)}(y,y';x),
\eeq
where
\bal
\label{3.26}
Q^{(n)}(y,y';x)=&\int_{-\frac{\epsilon}{2}}^{\frac{\epsilon}{2}}\dd y_1\int_{-\frac{\epsilon}{2}}^{\frac{\epsilon}{2}}\dd y_2\cdots\int_{-\frac{\epsilon}{2}}^{\frac{\epsilon}{2}}\dd y_{n-1}\nn
&\times R^{(n)}\left(x;\sum_{k=1}^n a(y_k-y_{k-1})\right),
\eal
with the convention $y_0\equiv y$, $y_n\equiv y'$, and with $R^{(n)}(x;\alpha)$ being defined by Eq.~\eqref{eq:qikj}.

The continuum version of Eq.~\eqref{3.5cc} is not straightforward. However, its equivalent form, Eq.~\eqref{B1cc}, becomes
\bal
\label{Fredh}
\frac{e^{-(s+\bp)a(y-y')}}{\lambda \phi(y)}=&\ell\frac{s+\bp}{\bp}\phi(y')\widetilde{G}(y,y';s)-\int_{-\frac{\epsilon}{2}}^{\frac{\epsilon}{2}}\dd y''\,\phi(y'')\nn
&\times \widetilde{G}(y,y'';s)e^{-(s+\bp)a(y''-y')}.
\eal
This is an inhomogeneous linear integral equation (of the second kind) for the Laplace transform, $\widetilde{G}(y,y';s)$, of ${g}(y,y';x)$.

As far as we know,  Eqs.~\eqref{3.25}, \eqref{3.26}, and \eqref{Fredh} had not been derived before.

\section{Results}
\label{sec:4}

 \subsection{The effect of finite $M$}\label{sec:result_finiteM}

Although we have expressed the results of Sec.~\ref{sec:3D} in the continuum limit, in practice we need to take a finite value of $M$ to obtain explicit results. We choose odd values of $M$ to include the centerline $y=0$ in the treatment.

 \begin{figure}
    \centering
    \includegraphics[width=0.95\columnwidth]{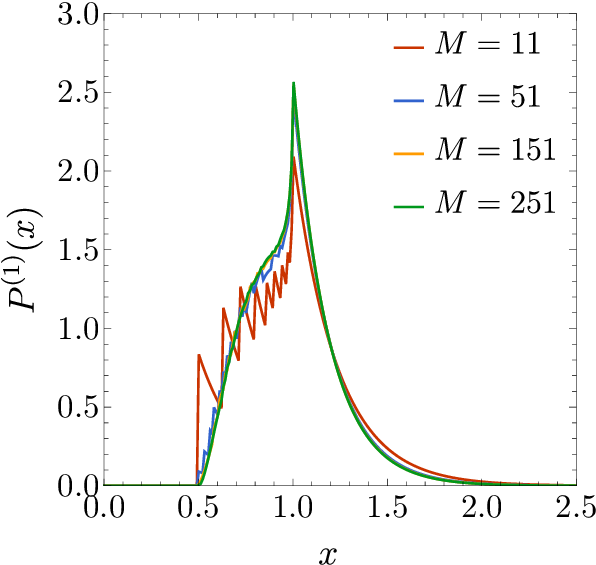}
    \caption{Nearest-neighbor probability distribution function $P^{(1)}(x)$ for a system with $\epsilon=\sqrt{3}/2$ at $\lambda=1.0$ and for different values of the discretization parameter $M$.}
    \label{fig:probabilitydensity_msize}
\end{figure}

Figure~\ref{fig:probabilitydensity_msize} shows the nearest-neighbor probability distribution function $P^{(1)}(x)$ for a system with the maximum pore width, $\epsilon=\epsilon_{\max}=\sqrt{3}/2\simeq 0.866$ (corresponding to $\lambda_\cl=2$), at a linear density $\lambda=1$ and for different values of $M$. We observe that the number of components $M=11$ is not large enough to capture satisfactorily well the expected form of $P^{(1)}(x)$ in the continuum. First, the discrete nature of the description is clearly apparent in the artificial jumps at $x=a\left(\frac{j-1}{M-1}\epsilon\right)$ with $j=2,\ldots,M$. Apart from that, the general shape of the function visibly deviates from the shape obtained with larger values of $M$. When taking $M=51$, the jumps at $x=a\left(\frac{j-1}{M-1}\epsilon\right)$ are much less pronounced and, moreover, the curve is rather close to that obtained with $M=151$ or $M=251$. Finally, the curves with the two latter values are practically indistinguishable from each other, which indicate a rapid convergence to the polydisperse limit.

In the remainder of the paper, all the presented calculations have been obtained with $M=251$,  unless explicitly stated otherwise.
An open-source C++ code used to procure the results of this section can be accessed from Ref.~\onlinecite{SingleFile2}.

\subsection{Neighbor probability distribution functions}

 \begin{figure}
    \includegraphics[width=0.95\columnwidth]{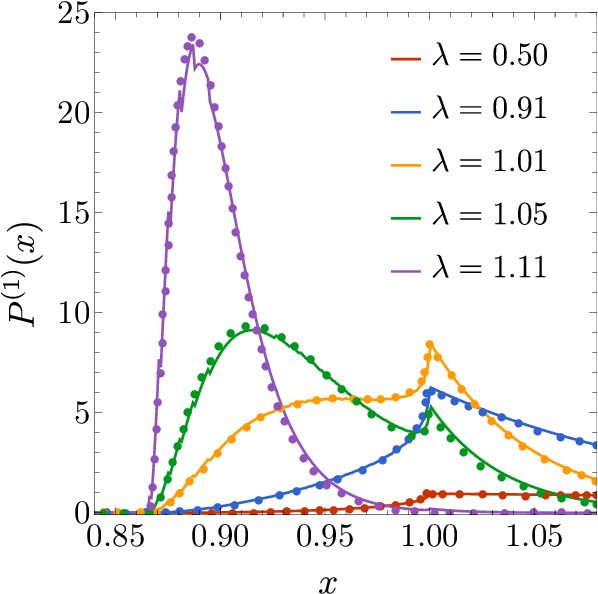}
    \caption{Nearest-neighbor probability distribution function $P^{(1)}(x)$ for a system with $\epsilon=\frac{1}{2}$ at several representative densities. Solid lines are our theoretical results, whereas symbols are MD data from Ref.~\onlinecite{HBPT21}.}
    \label{fig:probabilitydensity_nearest}
\end{figure}

Let us consider again the nearest-neighbor distribution $P^{(1)}(x)$. It is plotted in Fig.~\ref{fig:probabilitydensity_nearest} for $\epsilon=\frac{1}{2}$ (corresponding to $\lambda_\cl=1.1547$) and several  densities. An excellent agreement with molecular dynamics (MD) data\cite{HBPT21} is observed. Interestingly, as density decreases from values close to $\lambda_\cl$,  a secondary peak as a kink appears at $x\approx 1$. It becomes the main peak as density keeps decreasing; then, it is the only peak  and finally tends to soften for lower densities. The formation of this secondary peak was reported in Ref.~\onlinecite{HBPT20}, where it was shown to be related to the emergence of uncaging events in the zigzag-like array.

 \begin{figure}
    \includegraphics[width=0.95\columnwidth]{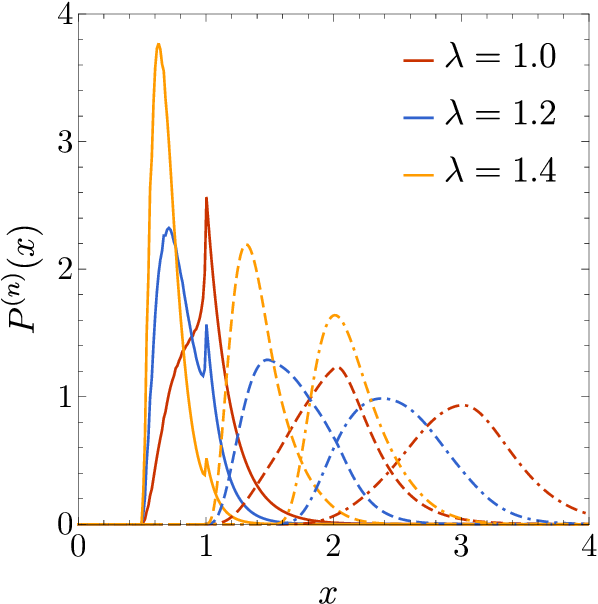}
    \caption{Probability distribution functions $P^{(n)}(x)$ with $n=1$ (solid lines), $2$ (dashed lines), and $3$ (dashed-dotted lines) for a system with $\epsilon=\sqrt{3}/2$ at different values of density.}
    \label{fig:probabilitydensity_n123}
\end{figure}

The $n$th neighbor probability distribution functions $P^{(n)}(x)$ with $n=1$, $2$, and $3$ are plotted in Fig.~\ref{fig:probabilitydensity_n123} for $\epsilon=\sqrt{3}/2$ and three densities.
As expected, $P^{(n)}(x)$ is nonzero only if $x>na(\epsilon)$. We also observe that $P^{(2)}(x)$ and $P^{(3)}(x)$ are much smoother than $P^{(1)}(x)$ and exhibit a single maximum. As density grows, the maximum moves toward $na(\epsilon)$ and becomes increasingly narrower.

 \subsection{Radial distribution functions}

\subsubsection{Total function}
After having studied the neighbor distributions $P^{(n)}(x)$, now, we turn to the RDF as the most relevant function. In our approach, $g(x)$ is analytically obtained from Eq.~\eqref{eq:gx_global} for $x\leq 3 a(\epsilon)$ (i.e., truncating the sum after $n=3$) and numerically from the Laplace inversion\cite{EulerILT} of Eq.~\eqref{3.5cc} for $x> 3 a(\epsilon)$. Notice that the planting method of Ref.~\onlinecite{HC21}, which  is essentially a numerical integration via random sampling, generates  alternative results to those of our numerical Laplace inversion.

\begin{figure}
    \includegraphics[width=0.95\columnwidth]{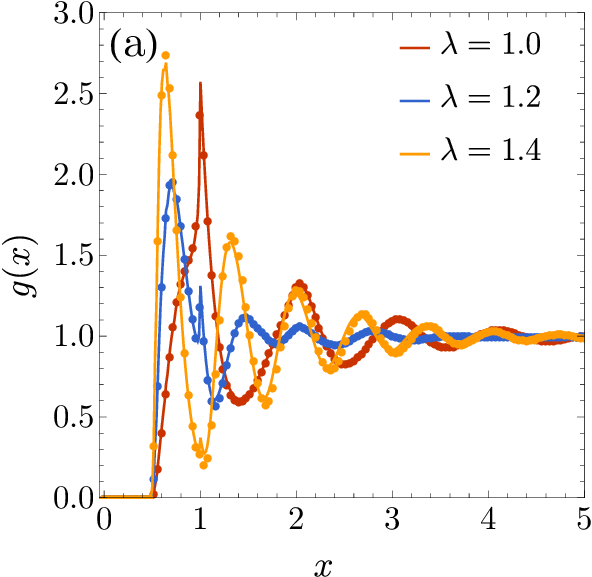}\\
    \includegraphics[width=0.95\columnwidth]{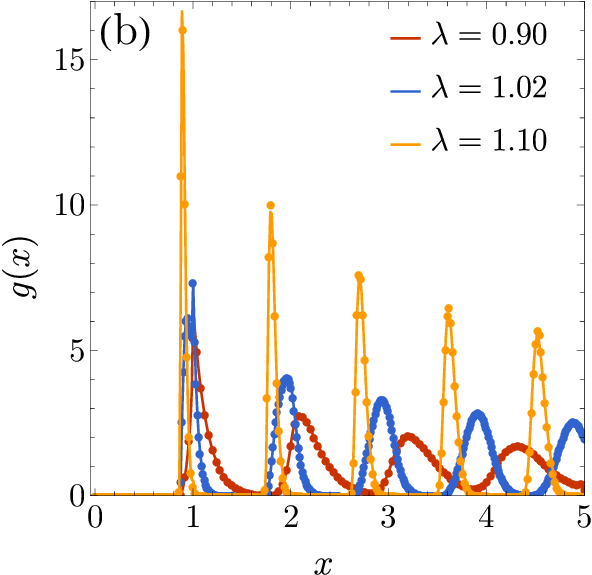}
    \caption{RDF $g(x)$ at different values of density for a system with (a) $\epsilon=\sqrt{3}/2$ and (b)  $\epsilon=\frac{1}{2}$. Solid lines are our theoretical results, whereas symbols are  MC data from Ref.~\onlinecite{VBG11}.}
    \label{fig:gr_e05}
\end{figure}

The results are illustrated in Fig.~\ref{fig:gr_e05}  for  $\epsilon=\sqrt{3}/2$  and $\epsilon=\frac{1}{2}$, in each case at three representative densities. The agreement with Monte Carlo (MC) simulation data\cite{VBG11} is very good.
Interesting structural features are observed in Fig.~\ref{fig:gr_e05}(a), where the densities are $50$\%--$70$\% of the close-packing value. As density increases, the structures become increasingly ordered, as illustrated by Fig.~\ref{fig:gr_e05}(b), where now the densities are $78$\%--$95$\% of the corresponding close-packing value.

\subsubsection{Partial functions}

In contrast to  $g(x)$, the partial RDF $g(y,y';x)$ describes spatial correlations between particles with \emph{specific}  transverse positions. Among all the possible choices of $y,y'$, the most interesting ones seem to be $\pm \frac{\epsilon}{2}$ and $0$. Thus, we focus on
\begin{subequations}
\label{4.1}
\beq
g_{++}(x)=g_{--}(x)\equiv g\left(\frac{\epsilon}{2},\frac{\epsilon}{2};x\right)=g\left(-\frac{\epsilon}{2},-\frac{\epsilon}{2};x\right),
\eeq
\beq
g_{+-}(x)=g_{-+}(x)\equiv g\left(\frac{\epsilon}{2},-\frac{\epsilon}{2};x\right)=g\left(-\frac{\epsilon}{2},\frac{\epsilon}{2};x\right),
\eeq
\beq
g_{00}(x)\equiv g\left(0,0;x\right),
\eeq
\beq
g_{+0}(x)=g_{-0}(x)\equiv g\left(\frac{\epsilon}{2},0;x\right)=g\left(-\frac{\epsilon}{2},0;x\right).
\eeq
\end{subequations}
The partial RDF $g_{++}(x)$ measures the longitudinal correlations between particles, both in contact with either the top or the bottom wall, whereas in the case of $g_{+-}(x)$ one of the particles is in contact with a wall and the other particle is in contact with the other wall. Similar interpretations can be assigned to $g_{00}(x)$ (both particles lie on the centerline) and   $g_{+0}(x)$ (one particle is in contact with a wall,  and the other one is on the centerline).

 \begin{figure*}
    \includegraphics[width=0.95\columnwidth]{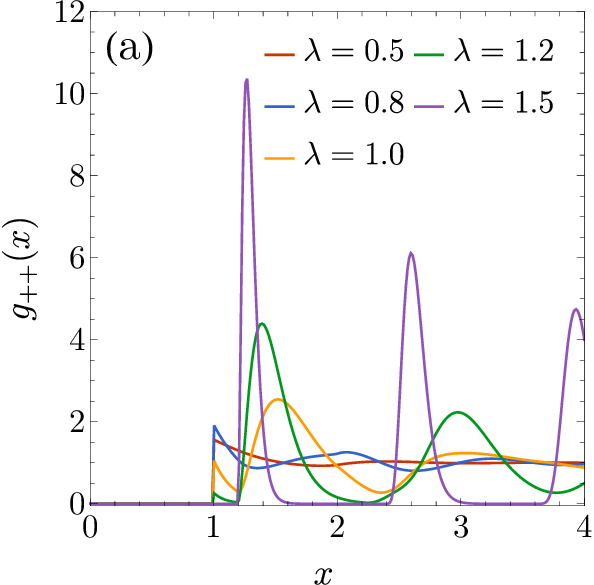}\includegraphics[width=0.95\columnwidth]{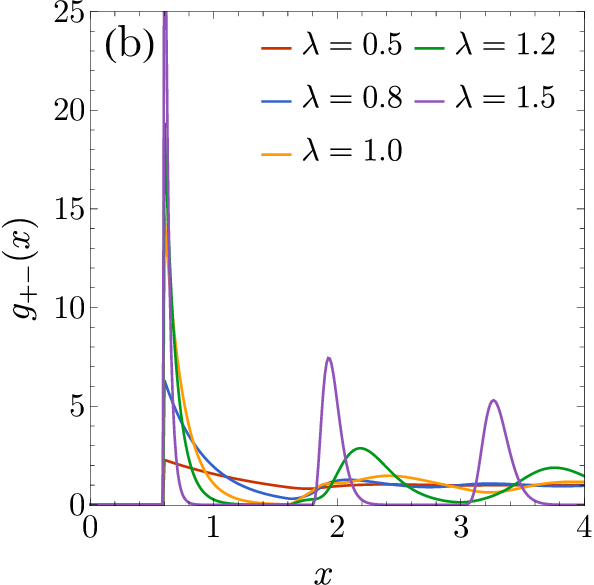}\\
    \includegraphics[width=0.95\columnwidth]{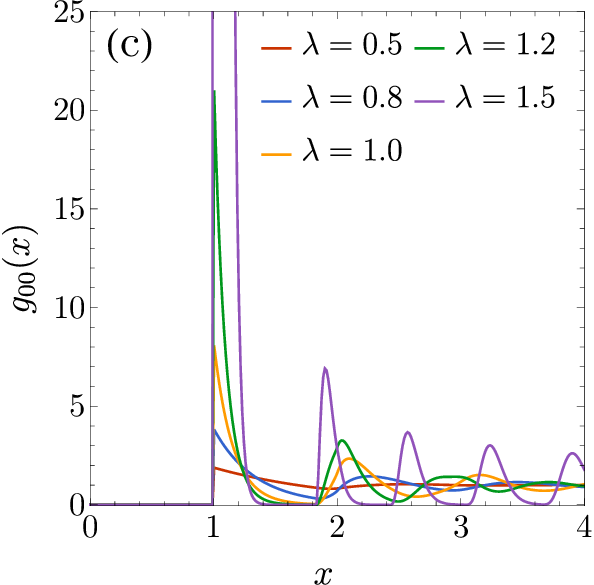}\includegraphics[width=0.95\columnwidth]{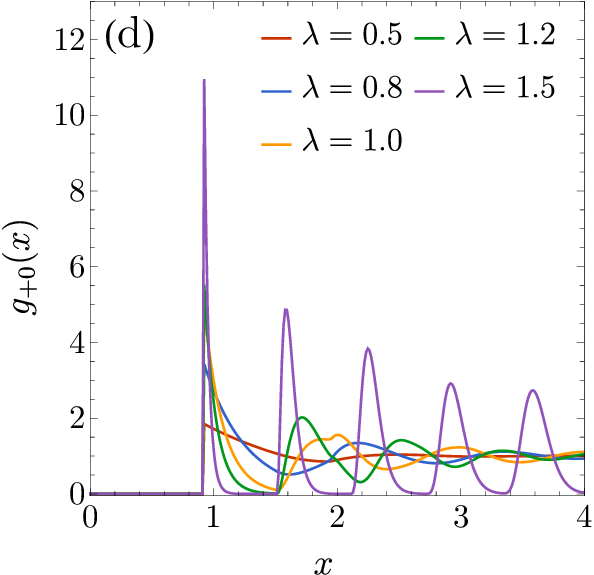}
    \caption{Plot of the partial RDF (a) $g_{++}(x)$, (b) $g_{+-}(x)$, (c) $g_{00}(x)$, and (d) $g_{+0}(x)$ for several densities and $\epsilon=0.8$.}
    \label{fig:gijr}
\end{figure*}

Figure~\ref{fig:gijr} shows the functions $g_{++}(x)$, $g_{+-}(x)$, $g_{00}(x)$, and $g_{+0}(x)$ for several densities and $\epsilon=0.8$, which corresponds to $\lambda_\cl\simeq 1.667$.
The contact distance is $x=1$ for both $g_{++}(x)$ and $g_{00}(x)$, but the contact value $g_{++}(1^+)$ is typically smaller than $g_{00}(1^+)$. The contact distances of $g_{+-}(x)$ and $g_{+0}(x)$ are $x=a(\epsilon)=0.6$ and $x=a(\frac{\epsilon}{2})\simeq 0.917$, respectively. As density increases, the contact value $g_{++}(1^+)$ starts growing, reaches a maximum, and then decreases.  Near close packing, $g_{++}(x)$ presents a depletion zone  between $x=1$ and $x=2a(\epsilon)$, together with pronounced peaks at $x\simeq 2a(\epsilon),4a(\epsilon), 6a(\epsilon), \ldots$. Also near close packing, the peaks of $g_{+-}(x)$, $g_{00}(x)$, and $g_{+0}(x)$ are located at $x\simeq a(\epsilon), 3a(\epsilon),5a(\epsilon),\ldots$, $x\simeq 1, 2a(\epsilon/2), 2a(\epsilon/2)+a(\epsilon),2a(\epsilon/2)+2a(\epsilon),2a(\epsilon/2)+3a(\epsilon),\ldots$, and $x\simeq a(\epsilon/2),a(\epsilon/2)+a(\epsilon),a(\epsilon/2)+2a(\epsilon),\ldots$, respectively. Note that the peak of $g_{00}(x)$ at $x=1^+$ for the density $\lambda=1.5$ is so high [$g_{00}(1^+)\simeq 4\times 10^3$] that it dramatically exceeds the vertical scale of Fig.~\ref{fig:gijr}(c).

\subsubsection{Disappearance of defects for high pressure}

   \begin{figure}
    \includegraphics[width=0.95\columnwidth]{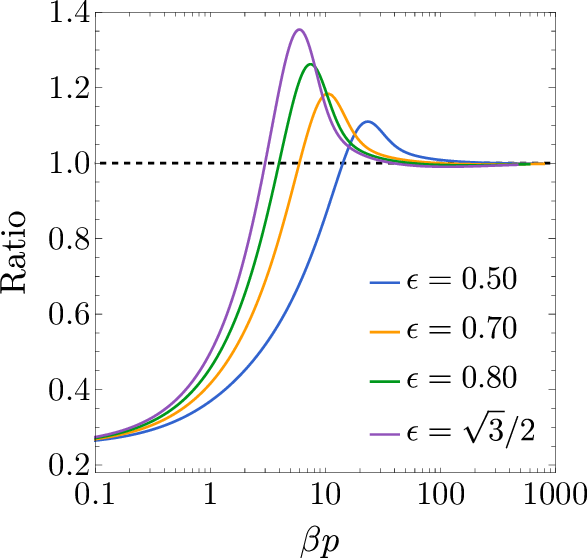}
    \caption{Ratio between the contact value $g_{++}(1^+)$ and its asymptotic form,  Eq.~\eqref{4.3a}, for several values of $\epsilon$.}
    \label{fig:ratio}
\end{figure}

All of this shows that a zigzag configuration ($\cdots+-+-+-+\cdots$) is clearly favored as the density approaches the close-packing value. On the other hand, this configuration  may present defects of the forms  $\cdots +-{++}-+-\cdots$ or  $\cdots +-+{--}+-\cdots$. This is quantified by a nonzero contact value $g_{++}(1^+)$, which decreases with increasing pressure. To study this effect in more detail, let us derive the high-pressure asymptotic behavior of $g_{++}(1^+)$. From Eq.~\eqref{3.24b}, one has
\beq
g_{++}(1^+)=\frac{Z}{\ell \phi^2(\frac{\epsilon}{2})}e^{-\bp}.
\eeq
In the high-pressure limit,\cite{MS23} $Z\to 2+\bp a(\epsilon)$, $\ell\to [a(\epsilon)/2\epsilon\bp] e^{-\bp a(\epsilon)}$, and $\phi(\frac{\epsilon}{2})\to\sqrt{\epsilon\bp /a(\epsilon)}$.
Therefore,
\beq
\label{4.3a}
g_{++}(1^+)\to 2[2+\bp a(\epsilon)]e^{-\bp[1-a(\epsilon)]}.
\eeq
Analogously, $g_{+-}(a(\epsilon)^+)=g_{++}(1^+)e^{\bp[1-a(\epsilon)]}\to 2[2+\bp a(\epsilon)]$.
Therefore, the defect quantifier $g_{++}(1^+)$ decays following the scaling form $g_{++}(1^+)\sim\bp e^{-\bp[1-a(\epsilon)]}$, while $g_{+-}(a(\epsilon)^+)$  increases linearly with pressure.
The ratio between $g_{++}(1^+)$ and its asymptotic form, as given by Eq.~\eqref{4.3a}, is plotted in Fig.~\ref{fig:ratio} for different values of $\epsilon$. We observe that higher pressures are needed to reach the asymptotic regime as the pore width decreases. This is because the different exponential terms in $g_{++}(1^+)$, which compete if $\bp\gg 1$, become more and more similar as $\epsilon$ decreases, and thus, the leading exponential needs increasingly higher pressures to dominate.

It might seem paradoxical that $g_{00}(1^+)$ diverges as density approaches close packing, although the population of particles at the centerline $y=0$ vanishes in that limit. However, we must recall that, as said at the end of Sec.~\ref{sec:systemDefinition}, the RDF $g(y,y';x)$ is the factor needed to get the two-body distribution $n_2(y,y';x)$ from the product $n_1(y)n_1(y')$, so that $n_2(0,0;1^+)=\lambda^2\phi^4(0)g_{00}(1^+)$. From the analysis in Ref.~\onlinecite{MS23}, one may estimate $\phi(0)\sim\sqrt{\bp}e^{-\bp[a(\frac{\epsilon}{2})-a(\epsilon)]}$ and $g_{00}(1^+)\sim \bp e^{\bp[2a(\frac{\epsilon}{2})-1-a(\epsilon)]}$, yielding $n_2(0,0;1^+)\sim (\bp)^{3} e^{-\bp[1+2a(\frac{\epsilon}{2})-3a(\epsilon)]}\to 0$, as expected.

\subsubsection{Asymptotic decay of the total correlation function. Correlation length and structural crossover}
The asymptotic decay of $h_{ij}(x)\equiv g_{ij}(x)-1$ is directly related to the nonzero poles $\{s_n\}$ of $\widetilde{G}_{ij}(s)$, i.e., the roots (different from $s=0$) of the determinant of the matrix $\mathsf{I}-A^2{\mathsf{\Omega}}(s+\bp)$ [see Eq.~\eqref{3.5cc}]. More explicitly,\cite{DE00,EHHPS93,FW69,PBH17,MS19}
\beq
\label{sec:fw1}
h_{ij}(x)= \sum_{n=1}^\infty \mathcal{A}_{ij,n} e^{s_n x},
\eeq
where the amplitudes $\mathcal{A}_{ij,n}=\mathrm{Res}\left[\widetilde{G}_{ij}(s)\right]_{s_n}$ are the associated residues. Although, in general, $\mathcal{A}_{ij,n}$ is different for each pair $ij$, the set of poles $\{s_n\}$ is common to all the pairs.
The {asymptotic} decay  of  $h_{ij}(x)$
is determined by the pair of conjugate poles, $s_\pm=-\kappa\pm\imath\omega$,  with the  real part closest to the origin, its residue being $|\mathcal{A}_{ij}|e^{\pm\imath\delta_{ij}}$. Therefore, for asymptotically large $x$,
\beq
\label{hij}
h_{ij}(x)\approx 2|\mathcal{A}_{ij}|e^{-\kappa x}\cos(\omega x+\delta_{ij}).
\eeq
As we see,  $\kappa^{-1}$ and $\omega/2\pi$ represent the longitudinal {correlation length} and the asymptotic oscillation frequency, respectively. As pressure increases, the damping coefficient decreases continuously. On the other hand, the angular frequency $\omega$ can experience a discontinuous jump at a certain pressure $\pc$, thus signaling a \emph{structural crossover} from oscillations with a given wavelength (if $p<\pc$) to oscillations with a different wavelength (if $p>\pc$).\cite{GDER04,GDER05,PYSHB21} This is due to a crossing of the real part of two competing poles with different imaginary parts.
Analogous crossovers in the transverse correlation length have been identified in Ref.~\onlinecite{HFC18} as  crossings in the two largest eigenvalues  of the transfer matrix.

\begin{figure}
    \includegraphics[width=0.95\columnwidth]{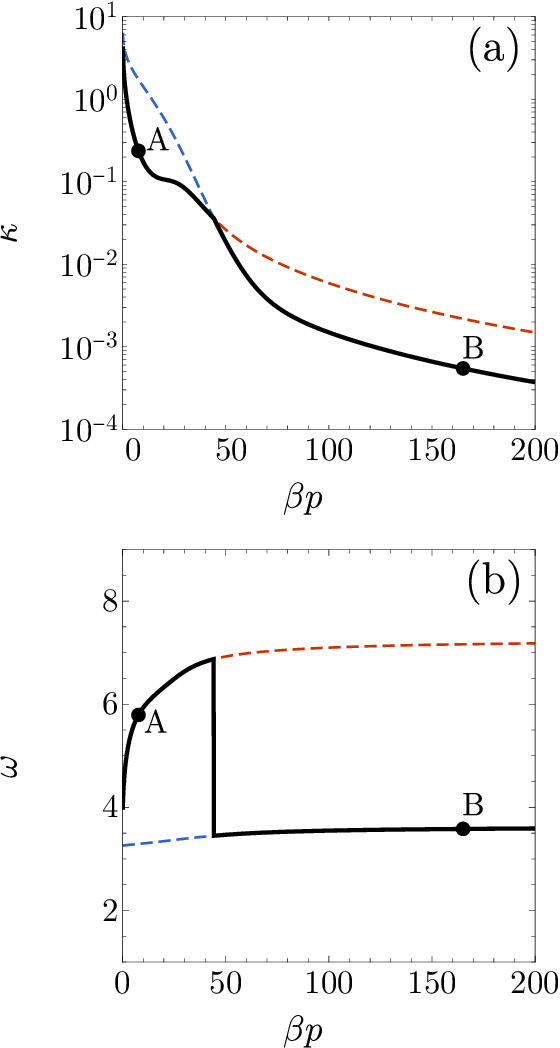}
    \caption{The thick solid lines represent (a) the damping coefficient $\kappa$ and (b) the frequency $\omega$   [see Eq.~\eqref{hij}], as functions of $\bp$, in the case $\epsilon=\frac{1}{2}$. In each panel, the dashed lines correspond to the continuation to $\bp>\bpc\simeq 44.2$ of the leading pole in the region $\bp<\bpc$, or vice versa. The circles with the labels A and B define the cases analyzed in Fig.~\ref{fig:asympt}.}
    \label{fig:poles}
\end{figure}

Taking a system with $\epsilon=\frac{1}{2}$ as an example, Fig.~\ref{fig:poles} shows the pressure dependence of both $\kappa$ and $\omega$. We observe that a structural crossover takes place at $\bpc\simeq 44.2$ (corresponding to $\lambdac\simeq 1.093$). For $p<\pc$, the oscillation wavelength ranges from $2\pi/\omega\simeq 1.57$ for low pressure to $2\pi/\omega\simeq 0.91$ near $\pc$, whereas it jumps to $2\pi/\omega\simeq 2a(\epsilon)= 1.732$ if $p>\pc$. This implies that for $p>\pc$ (or, equivalently, $\lambda>\lambdac$), the zigzag configuration persists for asymptotically large distances. According to the exponent in Eq.~\eqref{4.3a}, we can expect that $\bpc$ scales approximately with $1/[1-a(\epsilon)]$, thus decreasing with increasing $\epsilon$, as we have actually checked.
In what concerns the longitudinal correlation length $\kappa^{-1}$, it monotonically grows with pressure with a kink at $p=\pc$. In the high-pressure domain, we have checked that $\kappa^{-1}$ grows proportionally to $(\bp)^2$.

\begin{figure}
    \includegraphics[width=0.95\columnwidth]{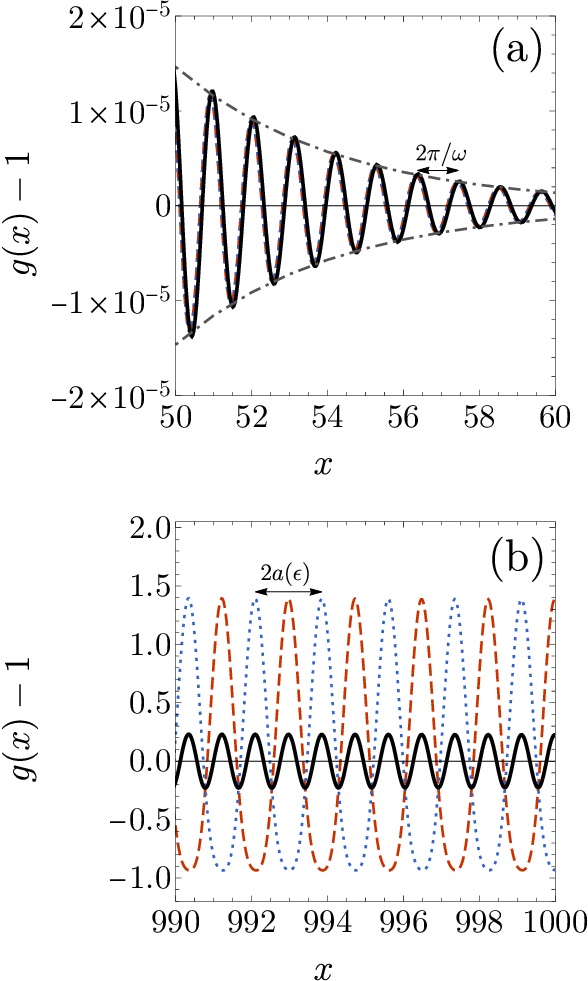}
    \caption{Large-$x$ behavior of the total correlation function $g(x)-1$ (solid lines) for $\epsilon=\frac{1}{2}$ and (a) $\lambda=0.9091$ ($\bp=7.68$, circle A in Fig.~\ref{fig:poles}) and (b) $\lambda=1.14$ ($\bp=165.1$, circle B in Fig.~\ref{fig:poles}). The dashed and dotted lines correspond to $g_{++}(x)-1$ and $g_{+-}(x)-1$, respectively. The dashed-dotted lines in panel (a) represent the exponential decay of the amplitudes, $\pm e^{-\kappa x}$ with $\kappa=0.237$. The horizontal double arrows indicate the wavelengths (a) $2\pi/\omega= 1.085$ and (b) $2\pi/\omega\simeq 2a(\epsilon)=1.732$.}
    \label{fig:asympt}
\end{figure}

To confirm the previous analysis, we have chosen the states A ($\bp=7.68$, $\lambda=0.9091$) and B ($\bp=165.1$, $\lambda=1.14$) as representative of cases with $p<\pc$ and $p>\pc$, respectively (see circles in Fig.~\ref{fig:poles}). For those states, $\kappa_A=0.237$, $\omega_A=5.792$, $\kappa_B=5.46\times 10^{-4}$, and $\omega_B=3.581$.
The results obtained from the numerical Laplace inversion for states A and B are plotted in Fig.~\ref{fig:asympt}. Apart from $h(x)$, the partial contributions $h_{++}(x)=g_{++}(x)-1$ and $h_{+-}(x)=g_{+-}(x)-1$ are also plotted. In state A ($p<\pc$), all the contributions $h_{ij}(x)$ oscillate in phase and practically with the same amplitude for large $x$. This explains why $h_{++}(x)$, $h_{+-}(x)$, and $h(x)$ are hardly distinguishable from each other in Fig.~\ref{fig:asympt}(a). In the case of state B ($p>\pc$), the amplitudes of $h_{++}(x)$ and $h_{+-}(x)$ keep being practically the same, but this time they are out-of-phase by a half-wavelength. This means that the short-distance shift $a(\epsilon)$ between $g_{+-}(x)$ and $g_{++}(x)$ [see Figs.~\ref{fig:gijr}(a) and \ref{fig:gijr}(b)] is maintained for large distances. As a consequence of this, the total function $h(x)$ oscillates with a smaller amplitude than $h_{++}(x)$ and $h_{+-}(x)$ and with a wavelength $a(\epsilon)$, which is the longitudinal distance between two adjacent disks in a zigzag configuration. It is interesting to note that the oscillations of $h_{++}(x)$ and $h_{+-}(x)$ in Fig.~\ref{fig:asympt}(a) are not purely harmonic since  hills are narrower and have a larger amplitude than the valleys. This indicates that the asymptotic decay with a single pole, Eq.~\eqref{hij}, has not been reached yet, as is also expected from the large amplitudes of $h_{++}(x)$ and $h_{+-}(x)$. However, the oscillations of the total function $h(x)\simeq \frac{1}{2}\left[h_{++}(x)+h_{+-}(x)\right]$ are almost perfectly harmonic.

\subsubsection{Two-dimensional radial distribution function}

\begin{figure}
    \includegraphics[width=0.95\columnwidth]{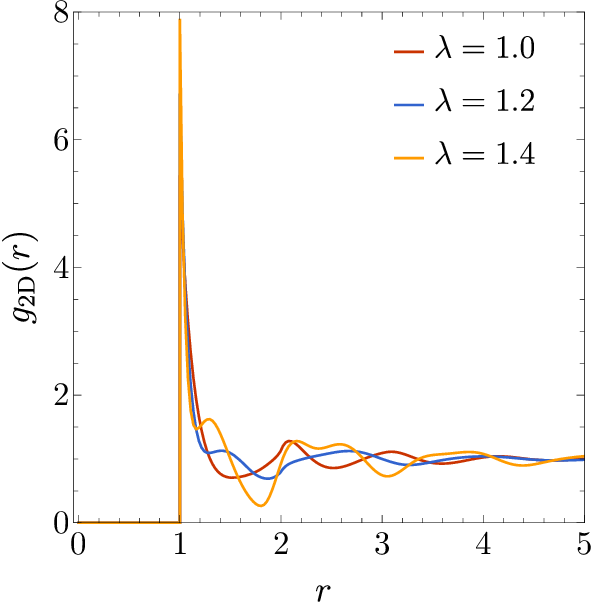}
   \caption{RDF $g_{\text{2D}}(r)$ for  a system with $\epsilon=\sqrt{3}/2$ at different values of density.}
    \label{fig:gr_2D}
\end{figure}

Now, we turn to the two-dimensional RDF $g_{\text{2D}}(r)$, defined by Eq.~\eqref{eq:g_2D_cont}. It is plotted in Fig.~\ref{fig:gr_2D} for a representative system with $\epsilon=\sqrt{3}/2$ and for the same values of $\lambda$ as in Figs.~\ref{fig:probabilitydensity_n123}  and \ref{fig:gr_e05}(a). Since this quantity is much more computationally demanding than $g(x)$ [compare Eqs.~\eqref{3.24c} and \eqref{eq:g_2D_cont}], we have taken $M=151$ in this case and checked that practical convergence is achieved with this value. We see from Fig.~\ref{fig:gr_2D}, the emergence of a secondary peak moving toward $2a(\epsilon)=1$ as density increases. Other interesting additional features are also observed.

 \subsection{Structure factor}

All the information contained in the RDF $g(x)$ is equivalently encapsulated in the static structure factor $S(q)$. Although the evaluation of the RDF for $x>3a(\epsilon)$ in our scheme is made by Laplace inversion of $\widetilde{G}(s)$, the structure factor is directly obtained from $\widetilde{G}(s)$ via Eq.~\eqref{eq:structure_factor}.
Alternatively, Robinson et al.\cite{RGM16} obtained $S(q)$ exactly from the transfer-matrix approach and used it to identify the onset of caging and the glassy behavior.

\begin{figure}
    \includegraphics[width=0.95\columnwidth]{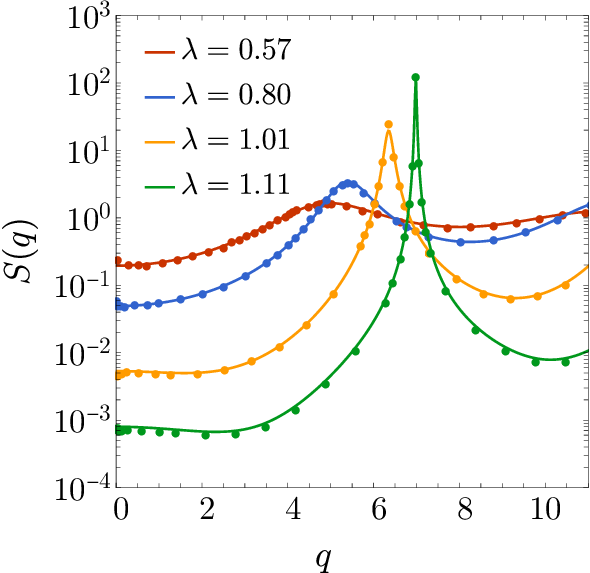}
    \caption{Structure factor $S(q)$ for a system with $\epsilon=\frac{1}{2}$ at several representative densities. Solid lines are our theoretical results, whereas symbols are MD data from Ref.~\onlinecite{HBPT21}.}
    \label{fig:structure_factor}
    \end{figure}

Figure~\ref{fig:structure_factor} shows $S(q)$ for $\epsilon=\frac{1}{2}$  and several  densities, with a very good agreement with MD data.\cite{HBPT21} As density approaches its close-packing value, $S(q)$ becomes more and more peaked around a density-dependent wave number $q_{\max}$. This signals an increasing ordering of the spatial correlations with a period $2\pi/q_{\max}$ slightly larger than the value $a(\epsilon)$ [see Figs.~\ref{fig:gr_e05}(b) and \ref{fig:asympt}(b)] associated with a zigzag pattern.

\begin{figure}
    \includegraphics[width=0.95\columnwidth]{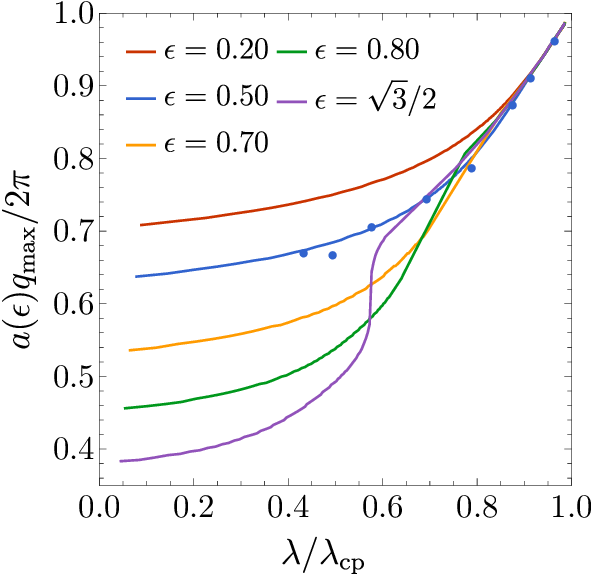}
    \caption{Scaled wave number $a(\epsilon)q_{\max}/2\pi$ vs the scaled density $\lambda/\lambda_\cl$ for several values of $\epsilon$. Solid lines are our theoretical results, whereas symbols for the case $\epsilon=\frac{1}{2}$ are MD data from Ref.~\onlinecite{HBPT21}.}
    \label{fig:structure_factor_kmax}
\end{figure}

The location of the first peak of $S(q)$, $q_{\max}$, is plotted in the scaled form in Fig.~\ref{fig:structure_factor_kmax} as a function of the scaled density $\lambda/\lambda_\cl$ for several values of the excess pore width $\epsilon$. In the case $\epsilon=\frac{1}{2}$, MD data from Ref.~\onlinecite{HBPT21} are also included, with a fair agreement, except for a small deviation at $\lambda=0.57$. As can be seen, and also observed in Fig.~\ref{fig:structure_factor}, the value of $q_{\max}$ increases with density, this effect being generally more pronounced as the pore width increases.
Interestingly,  the curve of  $q_{\max}$ vs $\lambda$ exhibits two inflection points if $\epsilon$ is high enough.

\section{Concluding Remarks}
\label{sec:conclusions}
In this work, we exploited the mapping of a Q1D hard-disk fluid onto a 1D non-additive mixture of hard rods with equal chemical potentials to obtain the (longitudinal) structural correlation functions of the original confined hard-disk fluid. Along the process, we first derived the exact thermodynamic properties (equation of state, Gibbs free energy, chemical potentials, and internal energy) for a generic 1D mixture with arbitrary number of components, $M$, arbitrary mole fractions, $\{x_i\}$, and arbitrary nearest-neighbor pair interactions, $\{\varphi_{ij}(x)\}$. Those thermodynamic quantities are expressed by Eqs.~\eqref{eq:number_density1}, \eqref{eq:Gibbs}, \eqref{eq:chemical_condition}, and \eqref{eq:internal_energy1}, where the dependence on temperature, pressure, and interaction potentials  occurs entirely through the parameters $\{A_i\}$ defined by the solution to Eq.~\eqref{eq:solve01}.

Particularization to our specific Q1D system requires the condition $A_i=A$, which fixes the mole fractions $\{x_i\to\phi^2(y_i)\delta y\}$, $\phi^2(y)$ representing the transverse density profile. Taking the continuum limit ($M\to\infty$), we were able to obtain an exact expression for the (partial) $n$th neighbor probability distribution function, $P^{(n)}(y,y';x)$, as given by Eqs.~\eqref{eq:qikj}, \eqref{3.25}, and \eqref{3.26}. From its knowledge, the total $n$th neighbor distribution, the partial RDF, and the total RDF can be obtained from Eqs.~\eqref{3.24a}, \eqref{3.24b}, and \eqref{3.24c}, respectively. Alternatively, the partial RDF is given in Laplace space as the solution of a linear integral equation, Eq.~\eqref{Fredh}.

From a practical point of view, the multiple $y$-integrals in Eqs.~\eqref{eq:z_exact_01}, \eqref{eq:2.5}, \eqref{3.24a}, and \eqref{3.26} need to be discretized for their evaluation, and this is equivalent to considering a discrete 1D mixture with a large number of components.
This discretization process is also essential to obtain the static structure factor $S(q)$ via Eqs.~\eqref{eq:structure_factor}, \eqref{3.5d}, and \eqref{3.5cc}. We showed that $M=251$ is sufficient to achieve convergence toward the continuum limit.

Explicit results for $P^{(n)}(x)$ (with $n=1,2,3$), $g(x)$, $g(y,y';x)$ (with $y,y'=0,\pm\epsilon/2$), and $S(q)$ were presented and discussed in Sec.~\ref{sec:4}. Comparison with available simulation data\cite{VBG11,HBPT21} showed an excellent agreement, thus validating the theoretical results derived in this paper, as well as the simulation techniques.

As an additional asset of our work, we have shown that the contact value $g_{++}(1^+)$, which can be interpreted as a signature of defects in the zigzag configuration, decays as $g_{++}(1^+)\sim \bp e^{-\bp[1-a(\epsilon)]}$ in the high-pressure limit. Interestingly, a structural crossover is found in the  frequency of the asymptotic oscillations of the RDF. Below a certain pressure ($\pc$), the oscillation wavelength decreases with increasing pressure. At $p=\pc$, a discontinuous jump to a larger wavelength close to $2a(\epsilon)$ occurs for $g(y,y';x)$, that wavelength becoming practically constant for $p>\pc$. Since in that high-pressure regime the oscillations of $g_{++}(x)$ and $g_{+-}(x)$ are out-of-phase by a distance $a(\epsilon)$, the wavelength of the oscillations of $g(x)$ turns out to be $a(\epsilon)$.

We hope that our research can stimulate the applications of the Q1D$\to$1D mapping to other systems. In particular, we plan to study the impact of a repulsive or attractive corona in the disks on the thermodynamic and structural properties of the confined fluid.

By using the same methodology, we also plan to study the case of hard spheres (of unit diameter) confined in a cylindrical pore of diameter $1+\epsilon$ with $\epsilon\leq \sqrt{3}/2$. In that system, the transverse position of a particle is given (in polar coordinates) by a vector $\mathbf{R}\equiv(R,\theta)$. Thus,  given two particles with transverse coordinates $\mathbf{R}$ and $\mathbf{R}'$, their longitudinal separation at contact is
$a(\mathbf{R},\mathbf{R}')=\sqrt{1-(R+R')^2+4R R'\cos^2\frac{\theta-\theta'}{2}}$.
Again, the original system can be mapped onto a  polydisperse and non-additive 1D mixture, where each component is identified by a vector $\mathbf{R}$ and the hard-core distance between particles belonging to species $\mathbf{R}$ and $\mathbf{R}'$ is $a(\mathbf{R},\mathbf{R}')$. In a discrete version of the mixture, each component is labeled by a pair $\mathbf{i}\equiv(i_R,i_\theta)$ with $i_R=0,1,\ldots,M_R$ and $i_\theta=1,2,\ldots, M_\theta$, so that $R_{i_R}=i_R \epsilon/2M_R$ and $\theta_{i_\theta}=(i_\theta-1)2\pi/M_\theta$. The expressions presented in Sec.~\ref{sec:3} keep being valid, except that $i\to\mathbf{i}$, $j\to\mathbf{j}$, and $a_{ij}\to a_{\mathbf{i}\mathbf{j}}=a(\mathbf{R}_{\mathbf{i}},\mathbf{R}_{\mathbf{j}})$, with $\mathbf{R}_{\mathbf{i}}\equiv (R_{i_R},\theta_{i_\theta})$.

\acknowledgments
We were grateful to the authors of Refs.~\onlinecite{VBG11,HBPT21} for providing us with the simulation data employed in Figs.~\ref{fig:probabilitydensity_nearest}, \ref{fig:gr_e05}, \ref{fig:structure_factor}, and \ref{fig:structure_factor_kmax}.
Financial support from Grant No.~PID2020-112936GB-I00 funded by the Spanish agency MCIN/AEI/10.13039/501100011033 and from Grant No.~IB20079 funded by Junta de Extremadura (Spain) and by ERDF
``A way of  making Europe'' was acknowledged.
A.M.M. was grateful to the Spanish Ministerio de Ciencia e Innovaci\'on for a predoctoral fellowship Grant No.~PRE2021-097702.

\section*{AUTHOR DECLARATIONS}
\subsection*{Conflict of Interest}
The authors have no conflicts to disclose.
\subsection*{Author Contributions}
\textbf{Ana M. Montero}: Formal analysis (equal); Investigation (equal);
Methodology (equal); Software (lead); Writing -- original draft
(lead). \textbf{Andr\'es Santos}: Conceptualization (lead); Formal analysis
(equal); Funding acquisition (lead); Investigation (equal); Methodology
(equal); Supervision (lead); Writing -- original draft (supporting);
Writing -- review \& editing (lead).

\section*{Data availability}
The data that support the findings of this study are available from the corresponding author upon reasonable request.

\appendix

\section{Isothermal susceptibility}
\label{app0}

It can be easily proved that $\partial_p\ell=-(\ell/p)(Z-1)$.\cite{MS23} Thus, Eq.~\eqref{eq:z_exact_01} yields
\bal
\label{compress1}
    \partial_p Z
    = &Z\frac{Z-1}{p}+\frac{\bp}{\ell} \int_{-\frac{\epsilon}{2}}^{\frac{\epsilon}{2}}\mathrm{d}y\int_{-\frac{\epsilon}{2}}^{\frac{\epsilon}{2}}\mathrm{d}y' \,e^{-\bp a(y-y')}a(y-y')\nn
    &\times\phi(y)\left[2\partial_p\phi(y')-\phi(y')\beta a(y-y')\right].
\eal
The isothermal susceptibility is $\chi_T=\beta^{-1}\partial_p\lambda=\partial_p(p/Z)$. Therefore, from Eq.~\eqref{compress1},
\bal
\label{compress2}
    \chi_T
    = &\frac{2-Z}{Z}-\frac{\lambda^2}{\beta\ell} \int_{-\frac{\epsilon}{2}}^{\frac{\epsilon}{2}}\mathrm{d}y\int_{-\frac{\epsilon}{2}}^{\frac{\epsilon}{2}}\mathrm{d}y' \,e^{-\bp a(y-y')}a(y-y')\nn
    &\times\phi(y)\left[2\partial_p\phi(y')-\phi(y')\beta a(y-y')\right].
\eal
It remains to determine the function $\partial_p\phi(y)$. Differentiating both sides of Eq.~\eqref{eq:2.5}, we obtain
\bal
\label{0.1}
\partial_p\phi(y)=&\frac{Z-1}{p}\phi(y)+\frac{1}{\ell}\int_{-\frac{\epsilon}{2}}^{\frac{\epsilon}{2}}\mathrm{d}y'\, e^{-\bp a(y-y')}\nn
 &\times\left[\partial_p\phi(y')-\phi(y')\beta a(y-y')\right].
\eal
This is an inhomogeneous linear integral equation (of the second kind) for $\partial_p\phi(y)$.

\section{Proof of Eq.~\eqref{eq:number_density}}
\label{appA}

Since $\lim_{x\to\infty}g_{ij}(x)=1$, the small-$s$ behavior of $\widetilde{G}_{ij}(s)$ must have the form $\widetilde{G}_{ij}(s)=s^{-1}+C_{ij}+\mathcal{O}(s)$. In the case of $\widetilde{P}_{ij}^{(1)}(s)$, Eq.~\eqref{3.5a} implies that $\widetilde{P}_{ij}^{(1)}(s)=\sqrt{x_j/x_i}A_i A_j\left[\Omega_{ij}(\bp)+\Omega_{ij}'(\bp)s+\mathcal{O}(s^2)\right]$, where $\Omega_{ij}'(s)\equiv \partial\Omega_{ij}(s)/\partial s$. Insertion of these expansions into Eq.~\eqref{B1} yields
\bal
\label{B2}
\frac{1}{\lambda}\sqrt{\frac{x_j}{x_i}}A_iA_j\Omega_{ij}(\bp)=&x_jC_{ij}-\sum_k\sqrt{x_j x_k}A_j A_k\nn
&\times\left[C_{ik}\Omega_{kj}(\bp)+\Omega_{kj}'(\bp)\right],
\eal
where use has been made of Eq.~\eqref{eq:solve01}. Summing over $j$ in both sides, and applying again Eq.~\eqref{eq:solve01}, we finally have
\beq
\label{B3}
\frac{1}{\lambda}=-\sum_{j,k}\sqrt{x_j x_k}A_j A_k\Omega_{kj}'(\bp).
\eeq
Since $\Omega_{kj}'(\bp)=\beta^{-1}\partial_p\Omega_{kj}(\bp)$, Eq.~\eqref{B3} becomes Eq.~\eqref{eq:number_density}.

\section{Proof of Eqs.~\eqref{3.19}--\eqref{eq:qikj}}\label{app:01}

Setting $\sqrt{x_i}=\phi_i$ and $A_i=A$ in Eq.~\eqref{3.5a}, and inserting the result in Eq.~\eqref{3.5b}, one finds
\beq
\label{A1}
\widetilde{P}_{ij}^{(n)}(s)=\frac{\phi_j}{\phi_i}A^{2n}\widetilde{Q}_{ij}^{(n)}(s),\quad \widetilde{\mathsf{Q}}^{(n)}(s)=\left[{\mathsf{\Omega}}(s+\bp)\right]^n,
\eeq
where the elements of the matrix $\mathsf{\Omega}(s)$ are $\Omega_{ij}(s)$.
More explicitly, the elements of the matrix $\widetilde{\mathsf{Q}}^{(n)}(s)$ are
\bal
\label{A2}
\widetilde{Q}_{ij}^{(n)}(s)=&\sum_{k_1}\sum_{k_2}\cdots\sum_{k_{n-1}}\Omega_{ik_1}(s+\bp)\Omega_{k_1k_2}(s+\bp)\cdots\nn
&\times \Omega_{k_{n-1}j}(s+\bp)\nn
=&\sum_{k_1}\sum_{k_2}\cdots\sum_{k_{n-1}}\widetilde{R}^{(n)}(s;a_{ik_1}+a_{k_1k_2}+\cdots+a_{k_{n-1}j}),
\eal
where
\beq
\label{A3}
\widetilde{R}^{(n)}(s;\alpha)\equiv \frac{e^{-(s+\bp)\alpha}}{(s+\bp)^n}.
\eeq
The inverse Laplace transform of $\widetilde{R}^{(n)}(s;\alpha)$ is given by Eq.~\eqref{eq:qikj}. Thus, Eqs.~\eqref{3.19} and \eqref{eq:qijn_expl} are readily obtained from Eqs.~\eqref{A1} and \eqref{A2}, respectively.

\section*{REFERENCES}



\begin{thebibliography}{56}%
\makeatletter
\providecommand \@ifxundefined [1]{%
 \@ifx{#1\undefined}
}%
\providecommand \@ifnum [1]{%
 \ifnum #1\expandafter \@firstoftwo
 \else \expandafter \@secondoftwo
 \fi
}%
\providecommand \@ifx [1]{%
 \ifx #1\expandafter \@firstoftwo
 \else \expandafter \@secondoftwo
 \fi
}%
\providecommand \natexlab [1]{#1}%
\providecommand \enquote  [1]{``#1''}%
\providecommand \bibnamefont  [1]{#1}%
\providecommand \bibfnamefont [1]{#1}%
\providecommand \citenamefont [1]{#1}%
\providecommand \href@noop [0]{\@secondoftwo}%
\providecommand \href [0]{\begingroup \@sanitize@url \@href}%
\providecommand \@href[1]{\@@startlink{#1}\@@href}%
\providecommand \@@href[1]{\endgroup#1\@@endlink}%
\providecommand \@sanitize@url [0]{\catcode `\\12\catcode `\$12\catcode
  `\&12\catcode `\#12\catcode `\^12\catcode `\_12\catcode `\%12\relax}%
\providecommand \@@startlink[1]{}%
\providecommand \@@endlink[0]{}%
\providecommand \url  [0]{\begingroup\@sanitize@url \@url }%
\providecommand \@url [1]{\endgroup\@href {#1}{\urlprefix }}%
\providecommand \urlprefix  [0]{URL }%
\providecommand \Eprint [0]{\href }%
\providecommand \doibase [0]{https://doi.org/}%
\providecommand \selectlanguage [0]{\@gobble}%
\providecommand \bibinfo  [0]{\@secondoftwo}%
\providecommand \bibfield  [0]{\@secondoftwo}%
\providecommand \translation [1]{[#1]}%
\providecommand \BibitemOpen [0]{}%
\providecommand \bibitemStop [0]{}%
\providecommand \bibitemNoStop [0]{.\EOS\space}%
\providecommand \EOS [0]{\spacefactor3000\relax}%
\providecommand \BibitemShut  [1]{\csname bibitem#1\endcsname}%
\let\auto@bib@innerbib\@empty
\bibitem [{\citenamefont {Barker}\ and\ \citenamefont
  {Henderson}(1976)}]{BH76}%
  \BibitemOpen
  \bibfield  {author} {\bibinfo {author} {\bibfnamefont {J.~A.}\ \bibnamefont
  {Barker}}\ and\ \bibinfo {author} {\bibfnamefont {D.}~\bibnamefont
  {Henderson}},\ }\bibfield  {title} {\enquote {\bibinfo {title} {What is
  ``liquid''? {Understanding} the states of matter},}\ }\href
  {https://doi.org/10.1103/RevModPhys.48.587} {\bibfield  {journal} {\bibinfo
  {journal} {Rev. Mod. Phys.}\ }\textbf {\bibinfo {volume} {48}},\ \bibinfo
  {pages} {587--671} (\bibinfo {year} {1976})}\BibitemShut {NoStop}%
\bibitem [{\citenamefont {Hansen}\ and\ \citenamefont {McDonald}(2013)}]{HM13}%
  \BibitemOpen
  \bibfield  {author} {\bibinfo {author} {\bibfnamefont {J.-P.}\ \bibnamefont
  {Hansen}}\ and\ \bibinfo {author} {\bibfnamefont {I.~R.}\ \bibnamefont
  {McDonald}},\ }\href@noop {} {\emph {\bibinfo {title} {{Theory of Simple
  Liquids}}}},\ \bibinfo {edition} {4th}\ ed.\ (\bibinfo  {publisher} {Academic
  Press},\ \bibinfo {address} {London},\ \bibinfo {year} {2013})\BibitemShut
  {NoStop}%
\bibitem [{\citenamefont {Santos}(2016)}]{S16}%
  \BibitemOpen
  \bibfield  {author} {\bibinfo {author} {\bibfnamefont {A.}~\bibnamefont
  {Santos}},\ }\href@noop {} {\emph {\bibinfo {title} {A Concise Course on the
  Theory of Classical Liquids. Basics and Selected Topics}}},\ \bibinfo
  {series} {Lecture Notes in Physics}, Vol.\ \bibinfo {volume} {923}\ (\bibinfo
   {publisher} {Springer},\ \bibinfo {address} {New York},\ \bibinfo {year}
  {2016})\BibitemShut {NoStop}%
\bibitem [{\citenamefont {Benavides}\ \emph {et~al.}(2006)\citenamefont
  {Benavides}, \citenamefont {del Pino}, \citenamefont {Gil-Villegas},\ and\
  \citenamefont {Sastre}}]{BPGS06}%
  \BibitemOpen
  \bibfield  {author} {\bibinfo {author} {\bibfnamefont {A.~L.}\ \bibnamefont
  {Benavides}}, \bibinfo {author} {\bibfnamefont {L.~A.}\ \bibnamefont {del
  Pino}}, \bibinfo {author} {\bibfnamefont {A.}~\bibnamefont {Gil-Villegas}},\
  and\ \bibinfo {author} {\bibfnamefont {F.}~\bibnamefont {Sastre}},\
  }\bibfield  {title} {\enquote {\bibinfo {title} {Thermodynamic and structural
  properties of confined discrete-potential fluids},}\ }\href
  {https://doi.org/10.1063/1.2382943} {\bibfield  {journal} {\bibinfo
  {journal} {J. Chem. Phys.}\ }\textbf {\bibinfo {volume} {125}},\ \bibinfo
  {pages} {204715} (\bibinfo {year} {2006})}\BibitemShut {NoStop}%
\bibitem [{\citenamefont {de~Oliveira}\ \emph {et~al.}(2006)\citenamefont
  {de~Oliveira}, \citenamefont {Netz}, \citenamefont {Colla},\ and\
  \citenamefont {Barbosa}}]{ONCB06}%
  \BibitemOpen
  \bibfield  {author} {\bibinfo {author} {\bibfnamefont {A.~B.}\ \bibnamefont
  {de~Oliveira}}, \bibinfo {author} {\bibfnamefont {P.~A.}\ \bibnamefont
  {Netz}}, \bibinfo {author} {\bibfnamefont {T.}~\bibnamefont {Colla}},\ and\
  \bibinfo {author} {\bibfnamefont {M.~C.}\ \bibnamefont {Barbosa}},\
  }\bibfield  {title} {\enquote {\bibinfo {title} {Structural anomalies for a
  three dimensional isotropic core-softened potential},}\ }\href
  {https://doi.org/10.1063/1.2357119} {\bibfield  {journal} {\bibinfo
  {journal} {J. Chem. Phys.}\ }\textbf {\bibinfo {volume} {125}},\ \bibinfo
  {pages} {124503} (\bibinfo {year} {2006})}\BibitemShut {NoStop}%
\bibitem [{\citenamefont {Robles}, \citenamefont {{L\'opez de Haro}},\ and\
  \citenamefont {Santos}(2007)}]{RHS07}%
  \BibitemOpen
  \bibfield  {author} {\bibinfo {author} {\bibfnamefont {M.}~\bibnamefont
  {Robles}}, \bibinfo {author} {\bibfnamefont {M.}~\bibnamefont {{L\'opez de
  Haro}}},\ and\ \bibinfo {author} {\bibfnamefont {A.}~\bibnamefont {Santos}},\
  }\bibfield  {title} {\enquote {\bibinfo {title} {{Percus--Yevick} theory for
  the structural properties of the seven-dimensional hard-sphere fluid},}\
  }\href {https://doi.org/10.1063/1.2424459} {\bibfield  {journal} {\bibinfo
  {journal} {J. Chem. Phys.}\ }\textbf {\bibinfo {volume} {126}},\ \bibinfo
  {pages} {{016}{101}} (\bibinfo {year} {2007})}\BibitemShut {NoStop}%
\bibitem [{\citenamefont {{Barraz Jr.}}, \citenamefont {Salcedo},\ and\
  \citenamefont {Barbosa}(2011)}]{BSB11}%
  \BibitemOpen
  \bibfield  {author} {\bibinfo {author} {\bibfnamefont {N.~M.}\ \bibnamefont
  {{Barraz Jr.}}}, \bibinfo {author} {\bibfnamefont {E.}~\bibnamefont
  {Salcedo}},\ and\ \bibinfo {author} {\bibfnamefont {M.~C.}\ \bibnamefont
  {Barbosa}},\ }\bibfield  {title} {\enquote {\bibinfo {title} {Thermodynamic,
  dynamic, structural, and excess entropy anomalies for core-softened
  potentials},}\ }\href {https://doi.org/10.1063/1.3630941} {\bibfield
  {journal} {\bibinfo  {journal} {J. Chem. Phys.}\ }\textbf {\bibinfo {volume}
  {135}},\ \bibinfo {pages} {104507} (\bibinfo {year} {2011})}\BibitemShut
  {NoStop}%
\bibitem [{\citenamefont {Muna\`o}\ and\ \citenamefont {Saija}(2016)}]{MS16}%
  \BibitemOpen
  \bibfield  {author} {\bibinfo {author} {\bibfnamefont {G.}~\bibnamefont
  {Muna\`o}}\ and\ \bibinfo {author} {\bibfnamefont {F.}~\bibnamefont
  {Saija}},\ }\bibfield  {title} {\enquote {\bibinfo {title} {Density and
  structural anomalies in soft-repulsive dimeric fluids},}\ }\href
  {https://doi.org/10.1039/c6cp00191b} {\bibfield  {journal} {\bibinfo
  {journal} {Phys. Chem. Chem. Phys.}\ }\textbf {\bibinfo {volume} {18}},\
  \bibinfo {pages} {9484--9489} (\bibinfo {year} {2016})}\BibitemShut {NoStop}%
\bibitem [{\citenamefont {Santos}, \citenamefont {Yuste},\ and\ \citenamefont
  {{L\'opez de Haro}}(2020)}]{SYH20}%
  \BibitemOpen
  \bibfield  {author} {\bibinfo {author} {\bibfnamefont {A.}~\bibnamefont
  {Santos}}, \bibinfo {author} {\bibfnamefont {S.~B.}\ \bibnamefont {Yuste}},\
  and\ \bibinfo {author} {\bibfnamefont {M.}~\bibnamefont {{L\'opez de
  Haro}}},\ }\bibfield  {title} {\enquote {\bibinfo {title} {Structural and
  thermodynamic properties of hard-sphere fluids},}\ }\href
  {https://doi.org/10.1063/5.0023903} {\bibfield  {journal} {\bibinfo
  {journal} {J. Chem. Phys.}\ }\textbf {\bibinfo {volume} {153}},\ \bibinfo
  {pages} {120901} (\bibinfo {year} {2020})}\BibitemShut {NoStop}%
\bibitem [{\citenamefont {Tonks}(1936)}]{T36}%
  \BibitemOpen
  \bibfield  {author} {\bibinfo {author} {\bibfnamefont {L.}~\bibnamefont
  {Tonks}},\ }\bibfield  {title} {\enquote {\bibinfo {title} {The complete
  equation of state of one, two and three-dimensional gases of hard elastic
  spheres},}\ }\href {https://doi.org/10.1103/PhysRev.50.955} {\bibfield
  {journal} {\bibinfo  {journal} {Phys. Rev.}\ }\textbf {\bibinfo {volume}
  {50}},\ \bibinfo {pages} {955--963} (\bibinfo {year} {1936})}\BibitemShut
  {NoStop}%
\bibitem [{\citenamefont {Katsura}\ and\ \citenamefont {Tago}(1968)}]{KT68}%
  \BibitemOpen
  \bibfield  {author} {\bibinfo {author} {\bibfnamefont {S.}~\bibnamefont
  {Katsura}}\ and\ \bibinfo {author} {\bibfnamefont {Y.}~\bibnamefont {Tago}},\
  }\bibfield  {title} {\enquote {\bibinfo {title} {Radial distribution function
  and the direct correlation function for one-dimensional gas with square-well
  potential},}\ }\href {https://doi.org/10.1063/1.1669764} {\bibfield
  {journal} {\bibinfo  {journal} {J. Chem. Phys.}\ }\textbf {\bibinfo {volume}
  {48}},\ \bibinfo {pages} {4246--4251} (\bibinfo {year} {1968})}\BibitemShut
  {NoStop}%
\bibitem [{\citenamefont {Heying}\ and\ \citenamefont {Corti}(2004)}]{HC04}%
  \BibitemOpen
  \bibfield  {author} {\bibinfo {author} {\bibfnamefont {M.}~\bibnamefont
  {Heying}}\ and\ \bibinfo {author} {\bibfnamefont {D.~S.}\ \bibnamefont
  {Corti}},\ }\bibfield  {title} {\enquote {\bibinfo {title} {The
  one-dimensional fully non-additive binary hard rod mixture: exact
  thermophysical properties},}\ }\href
  {https://doi.org/10.1016/j.fluid.2004.02.018} {\bibfield  {journal} {\bibinfo
   {journal} {Fluid Phase Equilib.}\ }\textbf {\bibinfo {volume} {220}},\
  \bibinfo {pages} {85--103} (\bibinfo {year} {2004})}\BibitemShut {NoStop}%
\bibitem [{\citenamefont {Schmidt}(2007)}]{S07b}%
  \BibitemOpen
  \bibfield  {author} {\bibinfo {author} {\bibfnamefont {M.}~\bibnamefont
  {Schmidt}},\ }\bibfield  {title} {\enquote {\bibinfo {title} {Fundamental
  measure density functional theory for nonadditive hard-core mixtures: {T}he
  one-dimensional case},}\ }\href {https://doi.org/10.1103/PhysRevE.76.031202}
  {\bibfield  {journal} {\bibinfo  {journal} {Phys. Rev. E}\ }\textbf {\bibinfo
  {volume} {76}},\ \bibinfo {pages} {{031}{202}} (\bibinfo {year}
  {2007})}\BibitemShut {NoStop}%
\bibitem [{\citenamefont {Santos}(2007)}]{S07}%
  \BibitemOpen
  \bibfield  {author} {\bibinfo {author} {\bibfnamefont {A.}~\bibnamefont
  {Santos}},\ }\bibfield  {title} {\enquote {\bibinfo {title} {Exact bulk
  correlation functions in one-dimensional nonadditive hard-core mixtures},}\
  }\href {https://doi.org/10.1103/PhysRevE.76.062201} {\bibfield  {journal}
  {\bibinfo  {journal} {Phys. Rev. E}\ }\textbf {\bibinfo {volume} {76}},\
  \bibinfo {pages} {{062}{201}} (\bibinfo {year} {2007})}\BibitemShut {NoStop}%
\bibitem [{\citenamefont {Varga}\ and\ \citenamefont {Gurin}(2011)}]{VG11}%
  \BibitemOpen
  \bibfield  {author} {\bibinfo {author} {\bibfnamefont {S.}~\bibnamefont
  {Varga}}\ and\ \bibinfo {author} {\bibfnamefont {P.}~\bibnamefont {Gurin}},\
  }\bibfield  {title} {\enquote {\bibinfo {title} {Towards understanding the
  ordering behavior of hard needles: {A}nalytical solutions in one
  dimension},}\ }\href {https://doi.org/10.1103/PhysRevE.83.061710} {\bibfield
  {journal} {\bibinfo  {journal} {Phys. Rev. E}\ }\textbf {\bibinfo {volume}
  {83}},\ \bibinfo {pages} {061710} (\bibinfo {year} {2011})}\BibitemShut
  {NoStop}%
\bibitem [{\citenamefont {Fantoni}\ and\ \citenamefont {Santos}(2017)}]{FS17}%
  \BibitemOpen
  \bibfield  {author} {\bibinfo {author} {\bibfnamefont {R.}~\bibnamefont
  {Fantoni}}\ and\ \bibinfo {author} {\bibfnamefont {A.}~\bibnamefont
  {Santos}},\ }\bibfield  {title} {\enquote {\bibinfo {title} {One-dimensional
  fluids with second nearest-neighbor interactions},}\ }\href
  {https://doi.org/10.1007/s10955-017-1908-6} {\bibfield  {journal} {\bibinfo
  {journal} {J. Stat. Phys.}\ }\textbf {\bibinfo {volume} {169}},\ \bibinfo
  {pages} {1171--1201} (\bibinfo {year} {2017})}\BibitemShut {NoStop}%
\bibitem [{\citenamefont {Montero}\ and\ \citenamefont {Santos}(2019)}]{MS19}%
  \BibitemOpen
  \bibfield  {author} {\bibinfo {author} {\bibfnamefont {A.~M.}\ \bibnamefont
  {Montero}}\ and\ \bibinfo {author} {\bibfnamefont {A.}~\bibnamefont
  {Santos}},\ }\bibfield  {title} {\enquote {\bibinfo {title} {Triangle-well
  and ramp interactions in one-dimensional fluids: A fully analytic exact
  solution},}\ }\href {https://doi.org/10.1007/s10955-019-02255-x} {\bibfield
  {journal} {\bibinfo  {journal} {J. Stat. Phys.}\ }\textbf {\bibinfo {volume}
  {175}},\ \bibinfo {pages} {269--288} (\bibinfo {year} {2019})}\BibitemShut
  {NoStop}%
\bibitem [{\citenamefont {Maestre}\ and\ \citenamefont {Santos}(2020)}]{MS20}%
  \BibitemOpen
  \bibfield  {author} {\bibinfo {author} {\bibfnamefont {M.~A.~G.}\
  \bibnamefont {Maestre}}\ and\ \bibinfo {author} {\bibfnamefont
  {A.}~\bibnamefont {Santos}},\ }\bibfield  {title} {\enquote {\bibinfo {title}
  {One-dimensional {J}anus fluids. {E}xact solution and mapping from the
  quenched to the annealed system},}\ }\href
  {https://doi.org/10.1088/1742-5468/ab900d} {\bibfield  {journal} {\bibinfo
  {journal} {J. Stat. Mech.}\ ,\ \bibinfo {pages} {063217}} (\bibinfo {year}
  {2020})}\BibitemShut {NoStop}%
\bibitem [{\citenamefont {Fantoni}, \citenamefont {Maestre},\ and\
  \citenamefont {Santos}()}]{FMS21}%
  \BibitemOpen
  \bibfield  {author} {\bibinfo {author} {\bibfnamefont {R.}~\bibnamefont
  {Fantoni}}, \bibinfo {author} {\bibfnamefont {M.~A.~G.}\ \bibnamefont
  {Maestre}},\ and\ \bibinfo {author} {\bibfnamefont {A.}~\bibnamefont
  {Santos}},\ }\bibfield  {title} {\enquote {\bibinfo {title} {Finite-size
  effects and thermodynamic limit in one-dimensional {J}anus fluids},}\ }\href
  {https://doi.org/10.1088/1742-5468/ac2897} {\bibfield  {journal} {\bibinfo
  {journal} {J. Stat. Mech,}\ }\textbf {\bibinfo {volume} {2021}},\ \bibinfo
  {pages} {103210}}\BibitemShut {NoStop}%
\bibitem [{\citenamefont {Barker}(1962)}]{B62}%
  \BibitemOpen
  \bibfield  {author} {\bibinfo {author} {\bibfnamefont {J.}~\bibnamefont
  {Barker}},\ }\bibfield  {title} {\enquote {\bibinfo {title} {Statistical
  mechanics of almost one-dimensional systems},}\ }\href
  {https://doi.org/10.1071/PH620127} {\bibfield  {journal} {\bibinfo  {journal}
  {Aust. J. Phys.,}\ }\textbf {\bibinfo {volume} {15}},\ \bibinfo {pages}
  {127--134} (\bibinfo {year} {1962})}\BibitemShut {NoStop}%
\bibitem [{\citenamefont {Barker}(1964)}]{B64b}%
  \BibitemOpen
  \bibfield  {author} {\bibinfo {author} {\bibfnamefont {J.}~\bibnamefont
  {Barker}},\ }\bibfield  {title} {\enquote {\bibinfo {title} {Statistical
  mechanics of almost one-dimensional systems. {II}},}\ }\href
  {https://doi.org/10.1071/PH640259} {\bibfield  {journal} {\bibinfo  {journal}
  {Aust. J. Phys.,}\ }\textbf {\bibinfo {volume} {17}},\ \bibinfo {pages}
  {259--268} (\bibinfo {year} {1964})}\BibitemShut {NoStop}%
\bibitem [{\citenamefont {Wojciechowski}, \citenamefont {Piera\'nski},\ and\
  \citenamefont {Ma{\l}ecki}(1982)}]{WPM82}%
  \BibitemOpen
  \bibfield  {author} {\bibinfo {author} {\bibfnamefont {K.~W.}\ \bibnamefont
  {Wojciechowski}}, \bibinfo {author} {\bibfnamefont {P.}~\bibnamefont
  {Piera\'nski}},\ and\ \bibinfo {author} {\bibfnamefont {J.}~\bibnamefont
  {Ma{\l}ecki}},\ }\bibfield  {title} {\enquote {\bibinfo {title} {A hard-disk
  system in a narrow box. {I}. {T}hermodynamic properties},}\ }\href
  {https://doi.org/10.1063/1.443019} {\bibfield  {journal} {\bibinfo  {journal}
  {J. Chem. Phys.}\ }\textbf {\bibinfo {volume} {76}},\ \bibinfo {pages}
  {6170--6175} (\bibinfo {year} {1982})}\BibitemShut {NoStop}%
\bibitem [{\citenamefont {Post}\ and\ \citenamefont {Kofke}(1992)}]{PK92}%
  \BibitemOpen
  \bibfield  {author} {\bibinfo {author} {\bibfnamefont {A.~J.}\ \bibnamefont
  {Post}}\ and\ \bibinfo {author} {\bibfnamefont {D.~A.}\ \bibnamefont
  {Kofke}},\ }\bibfield  {title} {\enquote {\bibinfo {title} {Fluids confined
  to narrow pores: A low-dimensional approach},}\ }\href
  {https://doi.org/10.1103/PhysRevA.45.939} {\bibfield  {journal} {\bibinfo
  {journal} {Phys. Rev. A}\ }\textbf {\bibinfo {volume} {45}},\ \bibinfo
  {pages} {939--952} (\bibinfo {year} {1992})}\BibitemShut {NoStop}%
\bibitem [{\citenamefont {Kofke}\ and\ \citenamefont {Post}(1993)}]{KP93}%
  \BibitemOpen
  \bibfield  {author} {\bibinfo {author} {\bibfnamefont {D.~A.}\ \bibnamefont
  {Kofke}}\ and\ \bibinfo {author} {\bibfnamefont {A.~J.}\ \bibnamefont
  {Post}},\ }\bibfield  {title} {\enquote {\bibinfo {title} {Hard particles in
  narrow pores. {T}ransfer-matrix solution and the periodic narrow box},}\
  }\href {https://doi.org/10.1063/1.464967} {\bibfield  {journal} {\bibinfo
  {journal} {J. Chem. Phys.}\ }\textbf {\bibinfo {volume} {98}},\ \bibinfo
  {pages} {4853--4861} (\bibinfo {year} {1993})}\BibitemShut {NoStop}%
\bibitem [{\citenamefont {Percus}(2002)}]{P02}%
  \BibitemOpen
  \bibfield  {author} {\bibinfo {author} {\bibfnamefont {J.~K.}\ \bibnamefont
  {Percus}},\ }\bibfield  {title} {\enquote {\bibinfo {title} {Density
  functional theory of single-file classical fluids},}\ }\href
  {https://doi.org/10.1080/00268970110109925} {\bibfield  {journal} {\bibinfo
  {journal} {Mol. Phys.}\ }\textbf {\bibinfo {volume} {100}},\ \bibinfo {pages}
  {2417--2422} (\bibinfo {year} {2002})}\BibitemShut {NoStop}%
\bibitem [{\citenamefont {Kamenetskiy}, \citenamefont {Mon},\ and\
  \citenamefont {Percus}(2004)}]{KMP04}%
  \BibitemOpen
  \bibfield  {author} {\bibinfo {author} {\bibfnamefont {I.~E.}\ \bibnamefont
  {Kamenetskiy}}, \bibinfo {author} {\bibfnamefont {K.~K.}\ \bibnamefont
  {Mon}},\ and\ \bibinfo {author} {\bibfnamefont {J.~K.}\ \bibnamefont
  {Percus}},\ }\bibfield  {title} {\enquote {\bibinfo {title} {Equation of
  state for hard-sphere fluid in restricted geometry},}\ }\href
  {https://doi.org/10.1063/1.1795131} {\bibfield  {journal} {\bibinfo
  {journal} {J. Chem. Phys.}\ }\textbf {\bibinfo {volume} {121}},\ \bibinfo
  {pages} {7355--7361} (\bibinfo {year} {2004})}\BibitemShut {NoStop}%
\bibitem [{\citenamefont {Forster}, \citenamefont {Mukamel},\ and\
  \citenamefont {Posch}(2004)}]{FMP04}%
  \BibitemOpen
  \bibfield  {author} {\bibinfo {author} {\bibfnamefont {C.}~\bibnamefont
  {Forster}}, \bibinfo {author} {\bibfnamefont {D.}~\bibnamefont {Mukamel}},\
  and\ \bibinfo {author} {\bibfnamefont {H.~A.}\ \bibnamefont {Posch}},\
  }\bibfield  {title} {\enquote {\bibinfo {title} {Hard disks in narrow
  channels},}\ }\href {https://doi.org/10.1103/PhysRevE.69.066124} {\bibfield
  {journal} {\bibinfo  {journal} {Phys. Rev. E}\ }\textbf {\bibinfo {volume}
  {69}},\ \bibinfo {pages} {066124} (\bibinfo {year} {2004})}\BibitemShut
  {NoStop}%
\bibitem [{\citenamefont {Varga}, \citenamefont {Ball{\'{o}}},\ and\
  \citenamefont {Gurin}()}]{VBG11}%
  \BibitemOpen
  \bibfield  {author} {\bibinfo {author} {\bibfnamefont {S.}~\bibnamefont
  {Varga}}, \bibinfo {author} {\bibfnamefont {G.}~\bibnamefont {Ball{\'{o}}}},\
  and\ \bibinfo {author} {\bibfnamefont {P.}~\bibnamefont {Gurin}},\ }\bibfield
   {title} {\enquote {\bibinfo {title} {Structural properties of hard disks in
  a narrow tube},}\ }\href {https://doi.org/10.1088/1742-5468/2011/11/p11006}
  {\bibfield  {journal} {\bibinfo  {journal} {J. Stat. Mech.}\ }\textbf
  {\bibinfo {volume} {2011}},\ \bibinfo {pages} {P11006}}\BibitemShut {NoStop}%
\bibitem [{\citenamefont {Gurin}\ and\ \citenamefont {Varga}(2013)}]{GV13}%
  \BibitemOpen
  \bibfield  {author} {\bibinfo {author} {\bibfnamefont {P.}~\bibnamefont
  {Gurin}}\ and\ \bibinfo {author} {\bibfnamefont {S.}~\bibnamefont {Varga}},\
  }\bibfield  {title} {\enquote {\bibinfo {title} {Pair correlation functions
  of two- and three-dimensional hard-core fluids confined into narrow pores:
  {E}xact results from transfer-matrix method},}\ }\href
  {https://doi.org/10.1063/1.4852181} {\bibfield  {journal} {\bibinfo
  {journal} {J. Chem. Phys.}\ }\textbf {\bibinfo {volume} {139}},\ \bibinfo
  {pages} {244708} (\bibinfo {year} {2013})}\BibitemShut {NoStop}%
\bibitem [{\citenamefont {Godfrey}\ and\ \citenamefont {Moore}(2014)}]{GM14}%
  \BibitemOpen
  \bibfield  {author} {\bibinfo {author} {\bibfnamefont {M.~J.}\ \bibnamefont
  {Godfrey}}\ and\ \bibinfo {author} {\bibfnamefont {M.~A.}\ \bibnamefont
  {Moore}},\ }\bibfield  {title} {\enquote {\bibinfo {title} {Static and
  dynamical properties of a hard-disk fluid confined to a narrow channel},}\
  }\href {https://doi.org/10.1103/PhysRevE.89.032111} {\bibfield  {journal}
  {\bibinfo  {journal} {Phys. Rev. E}\ }\textbf {\bibinfo {volume} {89}},\
  \bibinfo {pages} {032111} (\bibinfo {year} {2014})}\BibitemShut {NoStop}%
\bibitem [{\citenamefont {Mon}(2014)}]{M14b}%
  \BibitemOpen
  \bibfield  {author} {\bibinfo {author} {\bibfnamefont {K.~K.}\ \bibnamefont
  {Mon}},\ }\bibfield  {title} {\enquote {\bibinfo {title} {Third and fourth
  virial coefficients for hard disks in narrow channels},}\ }\href
  {https://doi.org/10.1063/1.4884607} {\bibfield  {journal} {\bibinfo
  {journal} {J. Chem. Phys.}\ }\textbf {\bibinfo {volume} {140}},\ \bibinfo
  {pages} {244504} (\bibinfo {year} {2014})}\BibitemShut {NoStop}%
\bibitem [{\citenamefont {Mon}(2015)}]{M15}%
  \BibitemOpen
  \bibfield  {author} {\bibinfo {author} {\bibfnamefont {K.~K.}\ \bibnamefont
  {Mon}},\ }\bibfield  {title} {\enquote {\bibinfo {title} {Erratum: `{T}hird
  and fourth virial coefficients for hard disks in narrow channels' [{J. Chem.
  Phys.} \textbf{140}, 244504 (2014)]},}\ }\href
  {https://doi.org/10.1063/1.4905470} {\bibfield  {journal} {\bibinfo
  {journal} {J. Chem. Phys.}\ }\textbf {\bibinfo {volume} {142}},\ \bibinfo
  {pages} {019901} (\bibinfo {year} {2015})}\BibitemShut {NoStop}%
\bibitem [{\citenamefont {Godfrey}\ and\ \citenamefont {Moore}(2015)}]{GM15}%
  \BibitemOpen
  \bibfield  {author} {\bibinfo {author} {\bibfnamefont {M.~J.}\ \bibnamefont
  {Godfrey}}\ and\ \bibinfo {author} {\bibfnamefont {M.~A.}\ \bibnamefont
  {Moore}},\ }\bibfield  {title} {\enquote {\bibinfo {title} {Understanding the
  ideal glass transition: {L}essons from an equilibrium study of hard disks in
  a channel},}\ }\href {https://doi.org/10.1103/PhysRevE.91.022120} {\bibfield
  {journal} {\bibinfo  {journal} {Phys. Rev. E}\ }\textbf {\bibinfo {volume}
  {91}},\ \bibinfo {pages} {022120} (\bibinfo {year} {2015})}\BibitemShut
  {NoStop}%
\bibitem [{\citenamefont {Hu}, \citenamefont {Fu},\ and\ \citenamefont
  {Charbonneau}(2018)}]{HFC18}%
  \BibitemOpen
  \bibfield  {author} {\bibinfo {author} {\bibfnamefont {Y.}~\bibnamefont
  {Hu}}, \bibinfo {author} {\bibfnamefont {L.}~\bibnamefont {Fu}},\ and\
  \bibinfo {author} {\bibfnamefont {P.}~\bibnamefont {Charbonneau}},\
  }\bibfield  {title} {\enquote {\bibinfo {title} {Correlation lengths in
  quasi-one-dimensional systems via transfer matrices},}\ }\href
  {https://doi.org/10.1080/00268976.2018.1479543} {\bibfield  {journal}
  {\bibinfo  {journal} {Mol. Phys.}\ }\textbf {\bibinfo {volume} {116}},\
  \bibinfo {pages} {3345--3354} (\bibinfo {year} {2018})}\BibitemShut {NoStop}%
\bibitem [{\citenamefont {Mon}(2020)}]{M20}%
  \BibitemOpen
  \bibfield  {author} {\bibinfo {author} {\bibfnamefont {K.~K.}\ \bibnamefont
  {Mon}},\ }\bibfield  {title} {\enquote {\bibinfo {title} {Analytical
  evaluation of third and fourth virial coefficients for hard disk fluids in
  narrow channels and equation of state},}\ }\href
  {https://doi.org/10.1016/j.physa.2020.124833} {\bibfield  {journal} {\bibinfo
   {journal} {Physica A}\ }\textbf {\bibinfo {volume} {556}},\ \bibinfo {pages}
  {124833} (\bibinfo {year} {2020})}\BibitemShut {NoStop}%
\bibitem [{\citenamefont {Huerta}\ \emph {et~al.}(2020)\citenamefont {Huerta},
  \citenamefont {Bryk}, \citenamefont {Pergamenshchik},\ and\ \citenamefont
  {Trokhymchuk}}]{HBPT20}%
  \BibitemOpen
  \bibfield  {author} {\bibinfo {author} {\bibfnamefont {A.}~\bibnamefont
  {Huerta}}, \bibinfo {author} {\bibfnamefont {T.}~\bibnamefont {Bryk}},
  \bibinfo {author} {\bibfnamefont {V.~M.}\ \bibnamefont {Pergamenshchik}},\
  and\ \bibinfo {author} {\bibfnamefont {A.}~\bibnamefont {Trokhymchuk}},\
  }\bibfield  {title} {\enquote {\bibinfo {title} {Kosterlitz-{T}houless-type
  caging-uncaging transition in a quasi-one-dimensional hard disk system},}\
  }\href {https://doi.org/10.1103/PhysRevResearch.2.033351} {\bibfield
  {journal} {\bibinfo  {journal} {Phys. Rev. Res.}\ }\textbf {\bibinfo {volume}
  {2}},\ \bibinfo {pages} {033351} (\bibinfo {year} {2020})}\BibitemShut
  {NoStop}%
\bibitem [{\citenamefont {Pergamenshchik}(2020)}]{P20}%
  \BibitemOpen
  \bibfield  {author} {\bibinfo {author} {\bibfnamefont {V.~M.}\ \bibnamefont
  {Pergamenshchik}},\ }\bibfield  {title} {\enquote {\bibinfo {title}
  {Analytical canonical partition function of a quasi-one-dimensional system of
  hard disks},}\ }\href {https://doi.org/10.1063/5.0025645} {\bibfield
  {journal} {\bibinfo  {journal} {J. Chem. Phys.}\ }\textbf {\bibinfo {volume}
  {153}},\ \bibinfo {pages} {144111} (\bibinfo {year} {2020})}\BibitemShut
  {NoStop}%
\bibitem [{\citenamefont {Pergamenshchik}, \citenamefont {Bryk},\ and\
  \citenamefont {Trokhymchuk}(2022)}]{PBT22}%
  \BibitemOpen
  \bibfield  {author} {\bibinfo {author} {\bibfnamefont {V.~M.}\ \bibnamefont
  {Pergamenshchik}}, \bibinfo {author} {\bibfnamefont {T.~M.}\ \bibnamefont
  {Bryk}},\ and\ \bibinfo {author} {\bibfnamefont {A.}~\bibnamefont
  {Trokhymchuk}},\ }\bibfield  {title} {\enquote {\bibinfo {title} {Correlation
  functions and ordering in a quasi-one dimensional system of hard disks from
  the exact canonical partition function},}\ }\href
  {https://doi.org/10.48550/arXiv.2206.05980} {\bibfield  {journal} {\bibinfo
  {journal} {arXiv:2206.05980}\ } (\bibinfo {year} {2022})}
  \BibitemShut {NoStop}%
\bibitem [{\citenamefont {Jung}\ and\ \citenamefont {Franosch}(2022)}]{JF22}%
  \BibitemOpen
  \bibfield  {author} {\bibinfo {author} {\bibfnamefont {G.}~\bibnamefont
  {Jung}}\ and\ \bibinfo {author} {\bibfnamefont {T.}~\bibnamefont
  {Franosch}},\ }\bibfield  {title} {\enquote {\bibinfo {title} {Structural
  properties of liquids in extreme confinement},}\ }\href
  {https://doi.org/10.1103/PhysRevE.106.014614} {\bibfield  {journal} {\bibinfo
   {journal} {Phys. Rev. E}\ }\textbf {\bibinfo {volume} {106}},\ \bibinfo
  {pages} {014614} (\bibinfo {year} {2022})}\BibitemShut {NoStop}%
\bibitem [{\citenamefont {Montero}\ and\ \citenamefont {Santos}(2023)}]{MS23}%
  \BibitemOpen
  \bibfield  {author} {\bibinfo {author} {\bibfnamefont {A.~M.}\ \bibnamefont
  {Montero}}\ and\ \bibinfo {author} {\bibfnamefont {A.}~\bibnamefont
  {Santos}},\ }\bibfield  {title} {\enquote {\bibinfo {title} {Equation of
  state of hard-disk fluids under single-file confinement},}\ }\href
  {https://doi.org/10.1063/5.0139116} {\bibfield  {journal} {\bibinfo
  {journal} {J. Chem. Phys.}\ }\textbf {\bibinfo {volume} {158}},\ \bibinfo
  {pages} {154501} (\bibinfo {year} {2023})}\BibitemShut {NoStop}%
\bibitem [{\citenamefont {Zhang}, \citenamefont {Godfrey},\ and\ \citenamefont
  {Moore}(2020)}]{ZGM20}%
  \BibitemOpen
  \bibfield  {author} {\bibinfo {author} {\bibfnamefont {Y.}~\bibnamefont
  {Zhang}}, \bibinfo {author} {\bibfnamefont {M.~J.}\ \bibnamefont {Godfrey}},\
  and\ \bibinfo {author} {\bibfnamefont {M.~A.}\ \bibnamefont {Moore}},\
  }\bibfield  {title} {\enquote {\bibinfo {title} {Marginally jammed states of
  hard disks in a one-dimensional channel},}\ }\href
  {https://doi.org/10.1103/PhysRevE.102.042614} {\bibfield  {journal} {\bibinfo
   {journal} {Phys. Rev. E}\ }\textbf {\bibinfo {volume} {102}},\ \bibinfo
  {pages} {042614} (\bibinfo {year} {2020})}\BibitemShut {NoStop}%
\bibitem [{\citenamefont {Huerta}\ \emph {et~al.}(2021)\citenamefont {Huerta},
  \citenamefont {Bryk}, \citenamefont {Pergamenshchik},\ and\ \citenamefont
  {Trokhymchuk}}]{HBPT21}%
  \BibitemOpen
  \bibfield  {author} {\bibinfo {author} {\bibfnamefont {A.}~\bibnamefont
  {Huerta}}, \bibinfo {author} {\bibfnamefont {T.}~\bibnamefont {Bryk}},
  \bibinfo {author} {\bibfnamefont {V.~M.}\ \bibnamefont {Pergamenshchik}},\
  and\ \bibinfo {author} {\bibfnamefont {A.}~\bibnamefont {Trokhymchuk}},\
  }\bibfield  {title} {\enquote {\bibinfo {title} {Collective dynamics in
  quasi-one-dimensional hard disk system},}\ }\href
  {https://doi.org/10.3389/fphy.2021.636052} {\bibfield  {journal} {\bibinfo
  {journal} {Front. Phys.}\ }\textbf {\bibinfo {volume} {9}},\ \bibinfo {pages}
  {636052} (\bibinfo {year} {2021})}\BibitemShut {NoStop}%
\bibitem [{\citenamefont {Hu}\ and\ \citenamefont {Charbonneau}(2021)}]{HC21}%
  \BibitemOpen
  \bibfield  {author} {\bibinfo {author} {\bibfnamefont {Y.}~\bibnamefont
  {Hu}}\ and\ \bibinfo {author} {\bibfnamefont {P.}~\bibnamefont
  {Charbonneau}},\ }\bibfield  {title} {\enquote {\bibinfo {title} {Comment on
  ``{K}osterlitz-{T}houless-type caging-uncaging transition in a
  quasi-one-dimensional hard disk system''},}\ }\href
  {https://doi.org/10.1103/PhysRevResearch.3.038001} {\bibfield  {journal}
  {\bibinfo  {journal} {Phys. Rev. Res.}\ }\textbf {\bibinfo {volume} {3}},\
  \bibinfo {pages} {038001} (\bibinfo {year} {2021})}\BibitemShut {NoStop}%
\bibitem [{\citenamefont {Robinson}, \citenamefont {Godfrey},\ and\
  \citenamefont {Moore}(2016)}]{RGM16}%
  \BibitemOpen
  \bibfield  {author} {\bibinfo {author} {\bibfnamefont {J.~F.}\ \bibnamefont
  {Robinson}}, \bibinfo {author} {\bibfnamefont {M.~J.}\ \bibnamefont
  {Godfrey}},\ and\ \bibinfo {author} {\bibfnamefont {M.~A.}\ \bibnamefont
  {Moore}},\ }\bibfield  {title} {\enquote {\bibinfo {title} {Glasslike
  behavior of a hard-disk fluid confined to a narrow channel},}\ }\href
  {https://doi.org/10.1103/PhysRevE.93.032101} {\bibfield  {journal} {\bibinfo
  {journal} {Phys. Rev. E}\ }\textbf {\bibinfo {volume} {93}},\ \bibinfo
  {pages} {032101} (\bibinfo {year} {2016})}\BibitemShut {NoStop}%
\bibitem [{\citenamefont {Banerjee}, \citenamefont {Jack},\ and\ \citenamefont
  {Cates}(2022)}]{BJC22}%
  \BibitemOpen
  \bibfield  {author} {\bibinfo {author} {\bibfnamefont {T.}~\bibnamefont
  {Banerjee}}, \bibinfo {author} {\bibfnamefont {R.~L.}\ \bibnamefont {Jack}},\
  and\ \bibinfo {author} {\bibfnamefont {M.~E.}\ \bibnamefont {Cates}},\
  }\bibfield  {title} {\enquote {\bibinfo {title} {Role of initial conditions
  in one-dimensional diffusive systems: {C}ompressibility, hyperuniformity, and
  long-term memory},}\ }\href {https://doi.org/10.1103/PhysRevE.106.L062101}
  {\bibfield  {journal} {\bibinfo  {journal} {Phys. Rev. E}\ }\textbf {\bibinfo
  {volume} {106}},\ \bibinfo {pages} {{L062101}} (\bibinfo {year}
  {2022})}\BibitemShut {NoStop}%
\bibitem [{not({\natexlab{a}})}]{note_23_04_1}%
  \BibitemOpen
  \href@noop {} {}  \bibinfo {note} {{N}ote that the quantity
  $K_{ij}$ defined in Chap.~5 of Ref.~\onlinecite{S16} is equivalent to
  $A_iA_j/\sqrt{x_i x_j}$.}\BibitemShut {Stop}%
\bibitem [{\citenamefont {Yuste}(2023)}]{EulerILT}%
  \BibitemOpen
  \bibfield  {author} {\bibinfo {author} {\bibfnamefont {S.~B.}\ \bibnamefont
  {Yuste}},\ }\href@noop {} {\enquote {\bibinfo {title} {{Numerical Inversion
  of Laplace Transforms using the Euler Method of Abate and Whitt}},}\
  }\bibinfo {howpublished}
  {\url{https://github.com/SantosBravo/Numerical-Inverse-Laplace-Transform-Abate-Whitt}}
  (\bibinfo {year} {2023}). \bibinfo {note} {Note that it is convenient to assign to the
  parameter \texttt{ntr} values larger than $x$.}\BibitemShut {Stop}%
\bibitem [{\citenamefont {Montero}(2023)}]{SingleFile2}%
  \BibitemOpen
  \bibfield  {author} {\bibinfo {author} {\bibfnamefont {A.~M.}\ \bibnamefont
  {Montero}},\ }\href@noop {} {\enquote {\bibinfo {title}
  {{SingleFileHardDisks-StructuralProperties}},}\ }\bibinfo {howpublished}
  {\url{https://github.com/amonterouex/SingleFileHardDisks-StructuralProperties}}
  (\bibinfo {year} {2023})\BibitemShut {NoStop}%
\bibitem [{\citenamefont {Dijkstra}\ and\ \citenamefont {Evans}(2000)}]{DE00}%
  \BibitemOpen
  \bibfield  {author} {\bibinfo {author} {\bibfnamefont {M.}~\bibnamefont
  {Dijkstra}}\ and\ \bibinfo {author} {\bibfnamefont {R.}~\bibnamefont
  {Evans}},\ }\bibfield  {title} {\enquote {\bibinfo {title} {A simulation
  study of the decay of the pair correlation function in simple fluids},}\
  }\href {https://doi.org/0.1063/1.480598} {\bibfield  {journal} {\bibinfo
  {journal} {J. Chem. Phys.}\ }\textbf {\bibinfo {volume} {112}},\ \bibinfo
  {pages} {1449--1456} (\bibinfo {year} {2000})}\BibitemShut {NoStop}%
\bibitem [{\citenamefont {Evans}\ \emph {et~al.}(1993)\citenamefont {Evans},
  \citenamefont {Henderson}, \citenamefont {Hoyle}, \citenamefont {Parry},\
  and\ \citenamefont {Sabeur}}]{EHHPS93}%
  \BibitemOpen
  \bibfield  {author} {\bibinfo {author} {\bibfnamefont {R.}~\bibnamefont
  {Evans}}, \bibinfo {author} {\bibfnamefont {J.~R.}\ \bibnamefont
  {Henderson}}, \bibinfo {author} {\bibfnamefont {D.~C.}\ \bibnamefont
  {Hoyle}}, \bibinfo {author} {\bibfnamefont {A.~O.}\ \bibnamefont {Parry}},\
  and\ \bibinfo {author} {\bibfnamefont {Z.~A.}\ \bibnamefont {Sabeur}},\
  }\bibfield  {title} {\enquote {\bibinfo {title} {Asymptotic decay of liquid
  structure: oscillatory liquid-vapour density profiles and the
  {F}isher--{W}idom line},}\ }\href {https://doi.org/10.1080/00268979300102621}
  {\bibfield  {journal} {\bibinfo  {journal} {Mol. Phys.}\ }\textbf {\bibinfo
  {volume} {80}},\ \bibinfo {pages} {755--775} (\bibinfo {year}
  {1993})}\BibitemShut {NoStop}%
\bibitem [{\citenamefont {Fisher}\ and\ \citenamefont {Widom}(1969)}]{FW69}%
  \BibitemOpen
  \bibfield  {author} {\bibinfo {author} {\bibfnamefont {M.~E.}\ \bibnamefont
  {Fisher}}\ and\ \bibinfo {author} {\bibfnamefont {B.}~\bibnamefont {Widom}},\
  }\bibfield  {title} {\enquote {\bibinfo {title} {Decay of correlations in
  linear systems},}\ }\href {https://doi.org/10.1063/1.1671624} {\bibfield
  {journal} {\bibinfo  {journal} {J. Chem. Phys.}\ }\textbf {\bibinfo {volume}
  {50}},\ \bibinfo {pages} {3756--3772} (\bibinfo {year} {1969})}\BibitemShut
  {NoStop}%
\bibitem [{\citenamefont {Pieprzyk}, \citenamefont {Bra\'{n}ka},\ and\
  \citenamefont {Heyes}(2017)}]{PBH17}%
  \BibitemOpen
  \bibfield  {author} {\bibinfo {author} {\bibfnamefont {S.}~\bibnamefont
  {Pieprzyk}}, \bibinfo {author} {\bibfnamefont {A.~C.}\ \bibnamefont
  {Bra\'{n}ka}},\ and\ \bibinfo {author} {\bibfnamefont {D.~M.}\ \bibnamefont
  {Heyes}},\ }\bibfield  {title} {\enquote {\bibinfo {title} {Representation of
  the direct correlation function of the hard-sphere fluid},}\ }\href
  {https://doi.org/10.1103/PhysRevE.95.062104} {\bibfield  {journal} {\bibinfo
  {journal} {Phys. Rev. E}\ }\textbf {\bibinfo {volume} {95}},\ \bibinfo
  {pages} {062104} (\bibinfo {year} {2017})}\BibitemShut {NoStop}%
\bibitem [{\citenamefont {Grodon}\ \emph {et~al.}(2004)\citenamefont {Grodon},
  \citenamefont {Dijkstra}, \citenamefont {Evans},\ and\ \citenamefont
  {Roth}}]{GDER04}%
  \BibitemOpen
  \bibfield  {author} {\bibinfo {author} {\bibfnamefont {C.}~\bibnamefont
  {Grodon}}, \bibinfo {author} {\bibfnamefont {M.}~\bibnamefont {Dijkstra}},
  \bibinfo {author} {\bibfnamefont {R.}~\bibnamefont {Evans}},\ and\ \bibinfo
  {author} {\bibfnamefont {R.}~\bibnamefont {Roth}},\ }\bibfield  {title}
  {\enquote {\bibinfo {title} {Decay of correlation functions in hard-sphere
  mixtures: Structural crossover},}\ }\href {https://doi.org/10.1063/1.1798057}
  {\bibfield  {journal} {\bibinfo  {journal} {J. Chem. Phys.}\ }\textbf
  {\bibinfo {volume} {121}},\ \bibinfo {pages} {7869--7882} (\bibinfo {year}
  {2004})}\BibitemShut {NoStop}%
\bibitem [{\citenamefont {Grodon}\ \emph {et~al.}(2005)\citenamefont {Grodon},
  \citenamefont {Dijkstra}, \citenamefont {Evans},\ and\ \citenamefont
  {Roth}}]{GDER05}%
  \BibitemOpen
  \bibfield  {author} {\bibinfo {author} {\bibfnamefont {C.}~\bibnamefont
  {Grodon}}, \bibinfo {author} {\bibfnamefont {M.}~\bibnamefont {Dijkstra}},
  \bibinfo {author} {\bibfnamefont {R.}~\bibnamefont {Evans}},\ and\ \bibinfo
  {author} {\bibfnamefont {R.}~\bibnamefont {Roth}},\ }\bibfield  {title}
  {\enquote {\bibinfo {title} {Homogeneous and inhomogeneous hard-sphere
  mixtures: manifestations of structural crossover},}\ }\href
  {https://doi.org/10.1080/00268970500167532} {\bibfield  {journal} {\bibinfo
  {journal} {Mol. Phys.}\ }\textbf {\bibinfo {volume} {103}},\ \bibinfo {pages}
  {3009--3023} (\bibinfo {year} {2005})}\BibitemShut {NoStop}%
\bibitem [{\citenamefont {Pieprzyk}\ \emph {et~al.}(2021)\citenamefont
  {Pieprzyk}, \citenamefont {Yuste}, \citenamefont {Santos}, \citenamefont
  {{L\'opez de Haro}},\ and\ \citenamefont {Bra\'nka}}]{PYSHB21}%
  \BibitemOpen
  \bibfield  {author} {\bibinfo {author} {\bibfnamefont {S.}~\bibnamefont
  {Pieprzyk}}, \bibinfo {author} {\bibfnamefont {S.~B.}\ \bibnamefont {Yuste}},
  \bibinfo {author} {\bibfnamefont {A.}~\bibnamefont {Santos}}, \bibinfo
  {author} {\bibfnamefont {M.}~\bibnamefont {{L\'opez de Haro}}},\ and\
  \bibinfo {author} {\bibfnamefont {A.~C.}\ \bibnamefont {Bra\'nka}},\
  }\bibfield  {title} {\enquote {\bibinfo {title} {Structural properties of
  additive binary hard-sphere mixtures. {II. A}symptotic behavior and
  structural crossovers},}\ }\href
  {https://doi.org/10.1103/PhysRevE.104.024128} {\bibfield  {journal} {\bibinfo
   {journal} {Phys. Rev. E}\ }\textbf {\bibinfo {volume} {104}},\ \bibinfo
  {pages} {024128} (\bibinfo {year} {2021})}\BibitemShut {NoStop}%
\end{thebibliography}

%

\end{document}